# Magnetic-field-induced delocalization in hybrid electron-nuclear spin ensembles

**Daniela Pagliero[1,*], Pablo R. Zangara[6,7,*], Jacob Henshaw[1], Ashok Ajoy[3], Rodolfo H. Acosta[6,7], Neil Manson[8], Jeffrey A. Reimer[4,5], Alexander Pines[3,5], Carlos A. Meriles[1,2,†]**

[1]*Department. of Physics, CUNY-City College of New York, New York, NY 10031, USA.*
[2]*CUNY-Graduate Center, New York, NY 10016, USA.*
[3]*Department of Chemistry, University of California at Berkeley, Berkeley, California 94720, USA.*
[4]*Department of Chemical and Biomolecular Engineering, University of California at Berkeley, Berkeley, California 94720, USA.*
[5]*Materials Science Division Lawrence Berkeley National Laboratory, Berkeley, California 94720, USA.*
[6]*Facultad de Matemática, Astronomía, Física y Computación, Universidad Nacional de Córdoba, Ciudad Universitaria, CP:X5000HUA Córdoba, Argentina.*
[7]*Instituto de Física Enrique Gaviola (IFEG), CONICET, Medina Allende s/n, X5000HUA, Córdoba, Argentina.*
[8]*Laser Physics Centre, Research School of Physics and Engineering, Australian National University, Canberra, A.C.T. 2601, Australia*

[*]Equally contributing authors. [†]Corresponding author. E-mail: cmeriles@ccny.cuny.edu

We use field-cycling-assisted dynamic nuclear polarization and continuous radio-frequency (RF) driving over a broad spectral range to demonstrate magnetic-field-dependent activation of nuclear spin transport from strongly-hyperfine-coupled $^{13}$C sites in diamond. We interpret our observations with the help of a theoretical framework where nuclear spin interactions are mediated by electron spins. In particular, we build on the results from a 4-spin toy model to show how otherwise localized nuclear spins must thermalize as they are brought in contact with a larger ancilla spin network. Further, by probing the system response to a variable driving field amplitude, we witness stark changes in the RF-absorption spectrum, which we interpret as partly due to contributions from heterogeneous multi-spin sets, whose 'zero-quantum' transitions become RF active thanks to the hybrid electron-nuclear nature of the system. These findings could prove relevant in applications to dynamic nuclear polarization, spin-based quantum information processing, and nanoscale sensing.

## I. INTRODUCTION

Nuclear spin-lattice relaxation in insulators is governed by interactions with paramagnetic centers within the material host, a notion first introduced by Bloembergen more than half a century ago[1]. Since these interactions strongly depend on the distance to the paramagnetic defect, the dynamics of nuclear spin thermalization emerges from an interplay between local relaxation rates and inter-nuclear couplings. In the simplest picture, nuclear spins sufficiently removed from the paramagnetic center converge jointly to a common temperature via spin diffusion, the energy-conserving process where a nuclear spin 'flips' at the expense of a 'flop' by a neighbor[2]. By contrast, strong magnetic field gradients near the defect — and the corresponding energy shifts they produce — disrupt spin exchange, prompting a description in terms of thermally disconnected regions of space — 'bulk' and 'local' spins — separated by a 'diffusion barrier'. The latter amounts to an imaginary surface where electron-nuclear and inter-nuclear spin couplings become comparable[3].

While the ideas above have undeniably proven valuable, they implicitly rest on a simplified scenario where the electronic spin bath is sufficiently dilute, i.e., where couplings between electronic spins are negligible. The impact these interactions can have in rendering the diffusion barrier permeable was first highlighted by Wolfe and collaborators in experiments with rare-earth-doped garnets at various concentrations[4,5]. More recently, the widespread use of dynamic nuclear polarization (DNP) methods have brought new attention to these early results as there is an inextricable connection between polarization flow and spin thermalization[6-9]. For example, experiments at low temperatures and high magnetic fields in radical-hosting organic matrices have exposed the combined impact of continuous microwave (MW) excitation and electron spectral diffusion on observed DNP 'spectra' (i.e., the observed nuclear magnetic resonance (NMR) signal as a function of the applied MW frequency)[10,11]. Further, electron-driven spin diffusion was introduced recently as a mechanism for nuclear polarization transfer in the proximity of paramagnetic defects[12]. Along related lines, DNP of carbon spins in diamond was exploited to reveal electron-spin-mediated nuclear spin diffusion exceeding the value expected for naturally abundant $^{13}$C spins by nearly two orders of magnitude[13].

Beyond applications to NMR signal enhancement, the interplay between diffusion and localization at the core of DNP can also be seen as an opportunity to investigate fundamental problems, most notably the competition between disorder and long-range interactions found in the out-of-equilibrium dynamics of driven open systems[14,15].

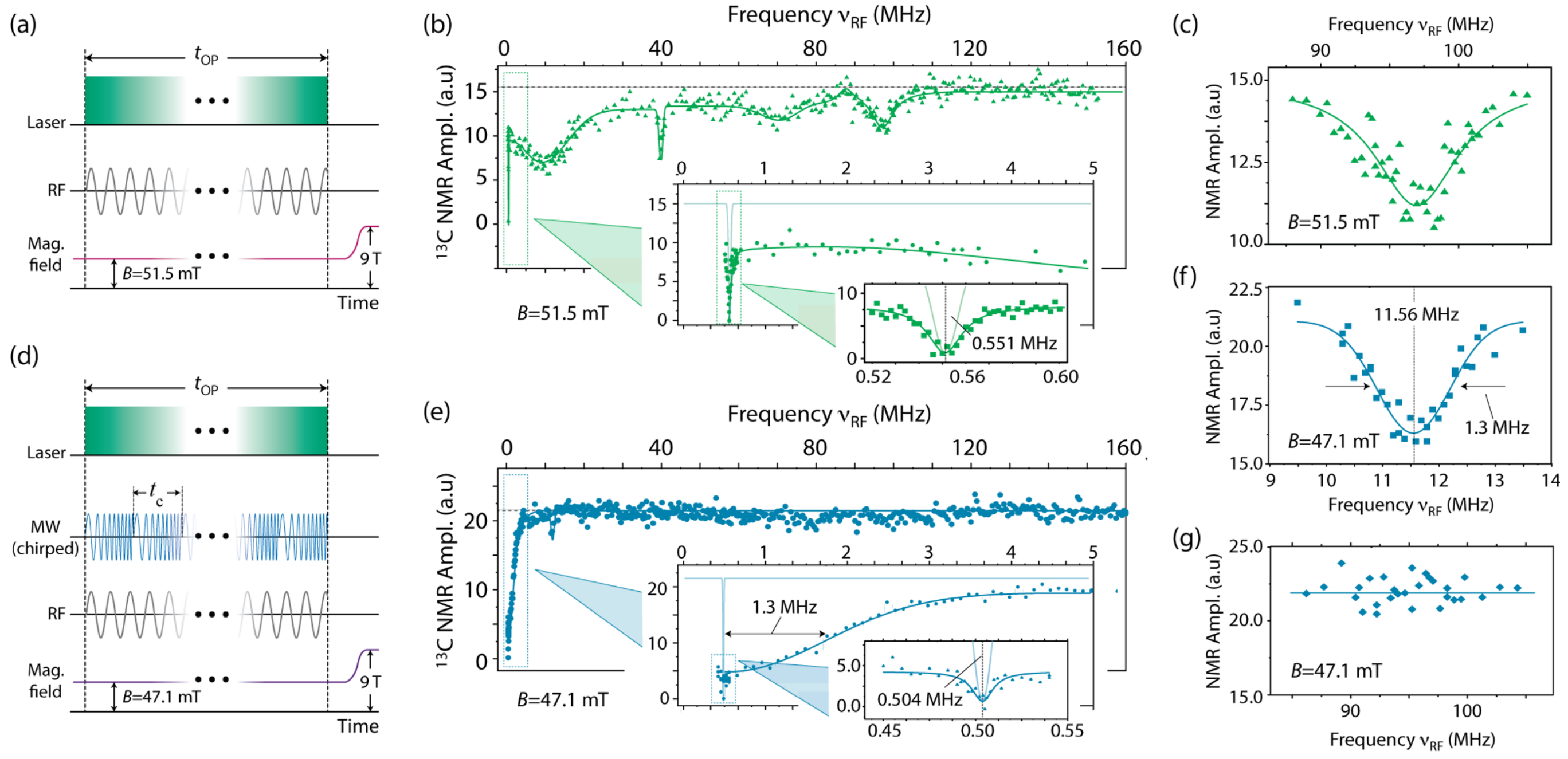

**Figure 1 | The role of P1 centers.** (a) Static matching field (SMF) protocol. (b) $^{13}$C NMR signal amplitude as a function of $\nu_{RF}$; the external magnetic field is $B = 51.5$ mT. (c) Zoomed SMF response around ~97 MHz. (d) Dynamic nuclear polarization via microwave sweeps (MWS). (e) Same as in (b) but using the MWS protocol to induce nuclear polarization; the external magnetic field is $B = 47.1$ mT. (f–g) Zoomed $^{13}$C response using the MWS protocol. Unlike (b), we see no high-frequency dips. In all experiments, $t_{OP} = 5$ s, the total number of repeats per point is 8, the driving field amplitude is $\Omega_{RF} = 4$ kHz, and the laser power is 1 W; solid traces are guides to the eye. In (d) through (g), the MW power is 300 mW, the sweep range is 25.2 MHz centered around the NV $|0\rangle \leftrightarrow |-1\rangle$ transition, the sweep rate is 15 MHz ms$^{-1}$ corresponding to a total of 8333 sweeps during $t_{OP}$.

Indeed, disorder and quantum interference can stymie thermalization, often leading to regimes of sub-diffusive dynamics or suppressed transport, a broad, fundamental phenomenon found in systems ranging from electrons in a crystal with disorder[16] to optical waves in a photonic structure[17]. Despite their differences, they all share similarities in that their Hamiltonians can often be mapped to those governing the dynamics of electron/nuclear spin sets in a solid.

Here, we resort to nuclear spins in diamond to demonstrate control over the localization/delocalization dynamics of hyperfine-coupled carbons upon variation of the applied magnetic field. We formally capture our observations by considering a model electron-nuclear spin chain featuring magnetic-field-dependent spin transport. Further, the dynamics at play can be cast in terms of distinct dynamic regimes that can be accessed by tuning the magnetic field strength and (effective) paramagnetic content. The spin state hybridization emerging from the intimate connection between electron and nuclear spins gives rise to otherwise forbidden low-frequency transitions, whose presence underlies the system's singular spectral response to RF excitation of variable amplitude.

## II. RESULTS

### A. Probing nuclear spin polarization transport at variable magnetic field

In our experiments, we dynamically polarize and probe $^{13}$C spins in a [100] diamond crystal (3×3×0.3 mm$^3$) grown in a high-pressure/high-temperature chamber (HPHT). The system is engineered to host a large (~10 ppm) concentration of nitrogen-vacancy (NV) centers, spin-1 paramagnetic defects that polarize efficiently under green illumination. Coexisting with the NVs is a more abundant group of P1 centers (~50 ppm), spin-1/2 defects formed by substitutional nitrogen atoms. We tune the externally applied magnetic field $B$ in and out the 'energy matching' range centered at $B_m$ (~51.8 mT for our present experimental conditions), where the Zeeman splitting of the P1 spins coincides with the frequency gap between the $|0\rangle$ and $|-1\rangle$ states of the NVs. Following electron and nuclear spin manipulation, we monitor the bulk $^{13}$C polarization via high-field NMR upon shuttling the sample into the bore of a 9 T magnet (additional experimental details can be found in Ref. (18)).

Fig. 1a shows a typical experimental protocol: We continuously illuminate the sample with a green laser (1 W at 532 nm) during a time interval $t_{OP} = 5$ s while simultaneously applying continuous radio-frequency (RF) excitation; here we set the field at 51.5 mT, slightly below $B_m$, where nuclear spins polarize positively as they provide the energy necessary to enable an NV-P1 'flip-flop'[18]. Figs. 1b and 1c show the resulting spectrum obtained as we measure the bulk $^{13}$C NMR signal for different RF frequencies $\nu_{RF}$ within the range 0.5-160 MHz. Besides the dip at 551 kHz — corresponding to the Larmor frequency of bulk $^{13}$C at $B = 51.5$ mT — we find several RF absorption bands, indicative of polarization transport from electron spins to bulk carbons via select groups of strongly hyperfine-coupled nuclei[13].

As an alternative to nuclear/electron spin cross-relaxation, one can dynamically polarize carbons via the use of chirped micro-wave (MW) pulses, consecutively applied during $t_{\mathrm{OP}}$[19,20] (Fig. 1d). Unlike the case above, nuclear spin polarization stems this time from Landau-Zener dynamics near level anti-crossings induced in the rotating frame as the MW sweeps the NV transitions[20] (specifically, the $|0\rangle \leftrightarrow |-1\rangle$ transition in the present case). Upon simultaneous RF excitation at variable frequencies, the spectrum that emerges indicates the polarization transport process is fundamentally distinct. This is shown in Figs. 1e through 1g, where we set the magnetic field to 47.1 mT, a shift of only ~4 mT from the experiments in Figs. 1a and 1b (yet sufficiently strong to quench cross-polarization-driven DNP[18]). In particular, we find that the RF impact is mostly limited to a ~1.3 MHz band adjacent to the $^{13}$C Larmor frequency (~0.5 MHz at 47 mT, insert in Fig. 1e). The differences are most striking near 40 MHz and 97 MHz where the dips observed at 51 mT (Figs. 1b and 1c) virtually vanish (Figs. 1e and 1g). Similarly, the small RF dip at ~11 MHz (Figs. 1e and 1f) amounts to only a little fraction of the broad absorption band centered at that frequency under field matching (Fig. 1b).

Before attempting to set these observations on a formal footing, we note that the generation and transport of nuclear spin polarization are two distinct physical processes: While the former provides the basis to understanding how order is transferred from electron to nuclear spins, our experiments allow us to investigate the latter, namely how strongly-hyperfine-coupled spins pass on polarization to 'bulk' nuclei (i.e., carbons whose hyperfine couplings are weaker than their mutual dipolar interactions). This question is particularly intriguing in diamond because $^{13}$C spins are relatively dilute (~1%) thus yielding weak dipolar couplings (~100 Hz), orders of magnitude smaller than typical hyperfine interactions (often in the ~1-10 MHz range and reaching up to ~130 MHz for first shell nuclei). Note that generation and transport are both necessary ingredients in the observation of DNP, implying that the absolute NMR signal amplitude per se — slightly different if cross-polarization or chirped MW is used to produce nuclear polarization, see Fig. 1 — has little intrinsic meaning. By contrast, we show below how the RF absorption spectra we measure allow us to gain a deeper understanding of the dynamics at play.

### B. Modeling transport via electron/nuclear spin sets

In the language of magnetic resonance, spin transport in DNP has been traditionally cast in terms of a 'spin-diffusion barrier', i.e., a virtual boundary around individual paramagnetic defects separating bulk spins from a 'frozen' nuclear core whose polarization cannot diffuse (simply because nuclear 'flip-flops' are energetically quenched). Avoiding such a scenario would require, in general, that polarization be generated via direct transfer from the defect to weakly coupled nuclei (featuring hyperfine constants of order ~100 Hz or less in the present case), a condition clearly inconsistent with the observations in Fig. 1 (both within or outside the NV/P1 field matching range). Further, the stark differences between the RF-absorption spectra observed in either case indicate that the very notion of a diffusion barrier as an inherent sample feature must be re-examined.

Although disorder in the crystal creates virtually countless combinations of interacting nuclear and electron spins, a concise description of nuclear spin transport demands the simplest possible spin set. On the other hand, the energy-conserving nature of this process imposes a minimum conceptual threshold: For instance, 3-spin sets — comprising, e.g., two electron spins and a carbon — provide an intuitive platform to describe polarization transfer from electrons to nuclei — the so-called 'cross effect' — but is clearly inadequate to describe polarization transport to bulk nuclei. Similar considerations apply to sets comprising two carbons and an electron spin because, under our experimental conditions, the energy change emerging from polarization hopping from one nuclear spin to the other is much smaller than the electron spin Zeeman energy at the applied magnetic field (~1.44 GHz), thus inhibiting electron/nuclear polarization transfer (see Section I in Ref. [21] for a formal discussion).

The above difficulties, however, can be circumvented with the toy model in Fig. 2a, a chain comprising an interacting pair of NV–P1 electron spins, each of them coupled to a neighboring carbon via hyperfine tensors of magnitude $\|A_j\|$ with $j = 1, 2$; for illustration purposes, we focus on the 'hyperfine-dominated' regime $\|A_1\| \sim \|A_2\| > \mathcal{J}_{\mathrm{d}} > \omega_{\mathrm{I}}$, where $\mathcal{J}_{\mathrm{d}}$ is the NV–P1 dipolar coupling constant, and $\omega_{\mathrm{I}}$ is the nuclear Larmor frequency. Intuitively, this system supports spin transport because changes in the nuclear and electronic spin energies compensate each other when the magnetic field takes on select transport-enabling values slightly shifted from $B_{\mathrm{m}}$, namely $B_{\mathrm{m}}^{(\varepsilon)} = B_{\mathrm{m}} + \delta B^{(\varepsilon)}$, with $\varepsilon = \alpha, \beta$, each corresponding to alternative sets of degenerate spin configurations of the chain[21].

In the absence of hyperfine couplings to the host nitrogen nucleus of either paramagnetic defect (a condition assumed here for simplicity), and using $\mathbf{I}_1$ ($\mathbf{I}_2$) to denote the vector spin operator of the nuclear spin coupled to the NV (P1), one can show that $^{13}$C spins in the chain are governed by the effective Hamiltonian[21]

$$H_{\mathrm{eff}} = \delta_{\mathrm{eff}} I_1^z - \delta_{\mathrm{eff}} I_2^z + J_{\mathrm{eff}}(I_1^+ I_2^- + I_1^- I_2^+), \qquad (1)$$

valid near either of the matching points. In the above expression, $\delta_{\mathrm{eff}} = 2\gamma_{\mathrm{e}} \left| B - B_{\mathrm{m}}^{(\alpha,\beta)} \right|$ is the effective nuclear spin frequency offset relative to the matching field $B_{\mathrm{m}}^{(\alpha,\beta)}$, $\gamma_{\mathrm{e}}$ is the electron spin gyromagnetic ratio, and we assume all spin operators are unit-less (i.e., $\hbar = 1$). Further, the effective coupling between nuclear spins is given by $J_{\mathrm{eff}} = -\omega_{\mathrm{I}} \mathcal{J}_d (A_2^{zx}/\Delta_2^2) \sin\left(\frac{\theta}{2}\right)$, where $\Delta_2^2 = (A_2^{zz})^2 + (A_2^{zx})^2$, $\tan(\theta) \approx A_1^{zx}/A_1^{zz}$, and $A_j^{zz}$ ($A_j^{zx}$) denotes the secular (pseudo-secular) hyperfine coupling constant for nuclear spin $j = 1, 2$.

Eq. (1) is a nuclear-spin-only Hamiltonian where paramagnetic interactions manifest in the form of field-dependent shifts and effective couplings largely exceeding the intrinsic $^{13}$C-$^{13}$C dipolar couplings. For example, for the present 50 ppm nitrogen concentration, we have $\mathcal{J}_{\mathrm{d}}\sim 3$ MHz and thus $J_{\mathrm{eff}}\sim 30$ kHz for $A_j^{zz}\sim A_j^{zx}\sim 10$ MHz, $j = 1, 2$. A numerical example demonstrating good agreement between the exact and effective nuclear spin evolution is presented in Fig. 2b for three different magnetic fields. It is worth highlighting the amplified sensitivity to field detuning $\left|B - B_{\mathrm{m}}^{(\alpha,\beta)}\right|$, impacting the offset terms in Eq. (1) via the electronic (not the nuclear) spin gyromagnetic ratio. We stress that the 4-spin model described above must be seen as the simplest set — among many others — compatible with an effective theory of nuclear magnetization transport as seen in our experiments. More general scenarios are discussed below.

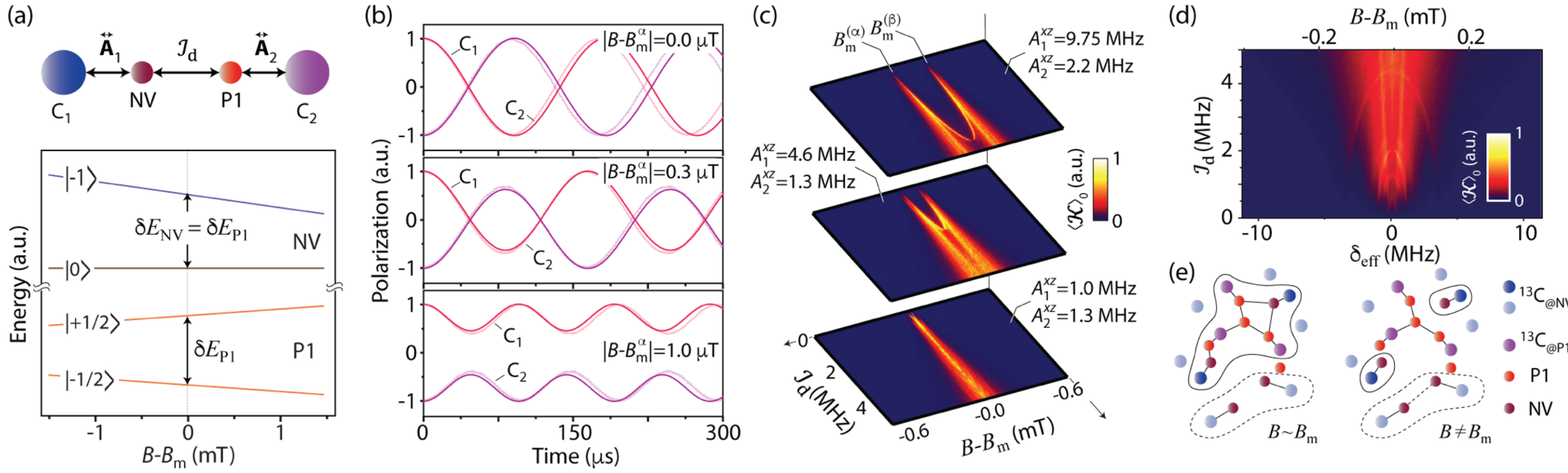

**Figure 2 | Magnetic-field-dependent spin transport.** (a) Model spin chain (top) and schematic NV–P1 energy diagram; at the matching field $B_{\mathrm{m}}$, the Zeeman splitting of the P1 coincides with the energy separation of the NV $|0\rangle \leftrightarrow |-1\rangle$ transition. (b) Inter-carbon polarization transfer for the chain in (a). The solid (faint) traces in each plot show the calculated evolution under the effective (exact) Hamiltonian assuming $\mathcal{J}_{\mathrm{d}} = 0.7$ MHz, $A_1^{xz} = 13$ MHz, $A_2^{xz} = 4$ MHz, and $A_j^{zz} = A_j^{zx}$ for $j = 1, 2$. (c) Nuclear spin current amplitude $\langle \mathcal{K} \rangle_0$ for the chain in (a) as a function of $B$ and $\mathcal{J}_{\mathrm{d}}$ for different hyperfine couplings. (d) Same as in (c) but after a weighted average over various configurations of hyperfine couplings (see Ref. 21). (e) Schematics of spin dynamics. (Left) When $B\sim B_{\mathrm{m}}$, $^{13}$C spins strongly coupled to NVs (dark blue and dark red circles, respectively) communicate with each other via networks formed by $^{13}$C spins hyperfine-coupled to P1s (purple and light red circles, respectively). (Right) Away from the energy matching range, strongly coupled carbons become localized. Weaker inter-NV interactions can mediate the transport of nuclear polarization seeded in carbons featuring intermediate or weak hyperfine couplings (light blue circles).

To more generally capture the nuclear spin dynamics prompted by NV–P1 couplings, we resort to the nuclear spin current operator $\mathcal{K} = (1/2i)(I_1^- I_2^+ - I_1^+ I_2^-)$, whose mean value — in general, a function of time $t$ — can be expressed as $\langle \mathcal{K} \rangle(t) = \langle \mathcal{K} \rangle_0 f(t)$, where $f(t)$ is a periodic function of unit amplitude[21]. Using $\langle \mathcal{K} \rangle_0$ as a measure of delocalization[22], we benchmark nuclear spin transport in Fig. 2c for different combinations of hyperfine couplings as a function of $B$ and $\mathcal{J}_{\mathrm{d}}$. We find non-zero transport within a confined region of the parameter space, with local maxima at fields $B_{\mathrm{m}}^{\alpha,\beta}$, discernible at weak inter-electronic couplings. Since these express the number of configurations compatible with nuclear spin transport, we anticipate additional matching fields should be present for more complex spin systems.

Our ability to externally activate transport is already implicit in the effective Hamiltonian in Eq. (1), which, upon the extension to a larger number of spins, can be mapped into the standard Anderson localization problem by means of the Wigner-Jordan transformation. Conceptually, the dynamics in the present spin system can be cast in terms of an interplay between 'disorder' — here expressed as site-selective nuclear Zeeman frequencies — and the amplitude of $^{13}$C–$^{13}$C 'flip-flop' couplings $J_{\mathrm{eff}}$ — also referred to as the 'hopping' term in charge transport studies. Sufficiently close to the matching condition, $\delta_{\mathrm{eff}} \lesssim J_{\mathrm{eff}}$ and the nuclear spins can flip-flop resonantly. On the other hand, a moderate detuning of the magnetic field yields $\delta_{\mathrm{eff}} \gg J_{\mathrm{eff}}$, putting the system back into a strongly localized dynamical phase. This is summarized in Fig. 2d where we compute a weighted average that takes into account the known set of carbon hyperfine couplings with the NV and P1 centers[21,23-26], and find non-zero current in the region where $J_{\mathrm{eff}} \gtrsim \delta_{\mathrm{eff}}$. We warn this latter condition must be understood in a distributional sense, i.e., for a given concentration of paramagnetic centers represented by $\mathcal{J}_{\mathrm{d}}$, there is a magnetic field range where spin diffusion channels become available to the most likely spin arrays in the crystal.

It is inevitable to draw a comparison between the distinct spin localization regimes we witness here and the dynamic phase diagram for charge carriers in a solid with disorder, as first introduced by Kimball[27]. Unfortunately, our experiments do not allow us to gradually transition from one regime to the other, with the consequence that we cannot presently probe criticality at the boundaries as seen in other experiments[28-31]. Assuming the proper experimental tools can be put in place, it will be interesting to devote additional work to characterize this system's response in intermediate regimes.

### C. Beyond the 4-spin model

Since the use of chirped MW pulses does produce observable $^{13}$C signal, it is clear that spin transport outside the above field range is also granted, though the observations in Fig. 1 indicate the enabling channels are different. Direct MW-assisted polarization of bulk nuclei can be ruled out immediately because the results in Fig. 1e show $^{13}$C spins with couplings as large as ~1 MHz — approximately 4 orders of magnitude greater than homonuclear interactions — do play a key role in the transport. Further, a detailed analysis of chirp-pulse-driven DNP in diamond[20] shows that polarization transfer — governed by Landau-Zener dynamics at level anti-crossings in the rotating frame — is highly efficient for carbons featuring hyperfine couplings greater than ~1 MHz, but decays sharply for more-weakly interacting nuclei. Correspondingly, the sharp differences between the RF absorption spectra in Fig. 1 point to a distinct polarization transport mechanism where strongly coupled carbons, though polarized, communicate with the rest less efficiently.

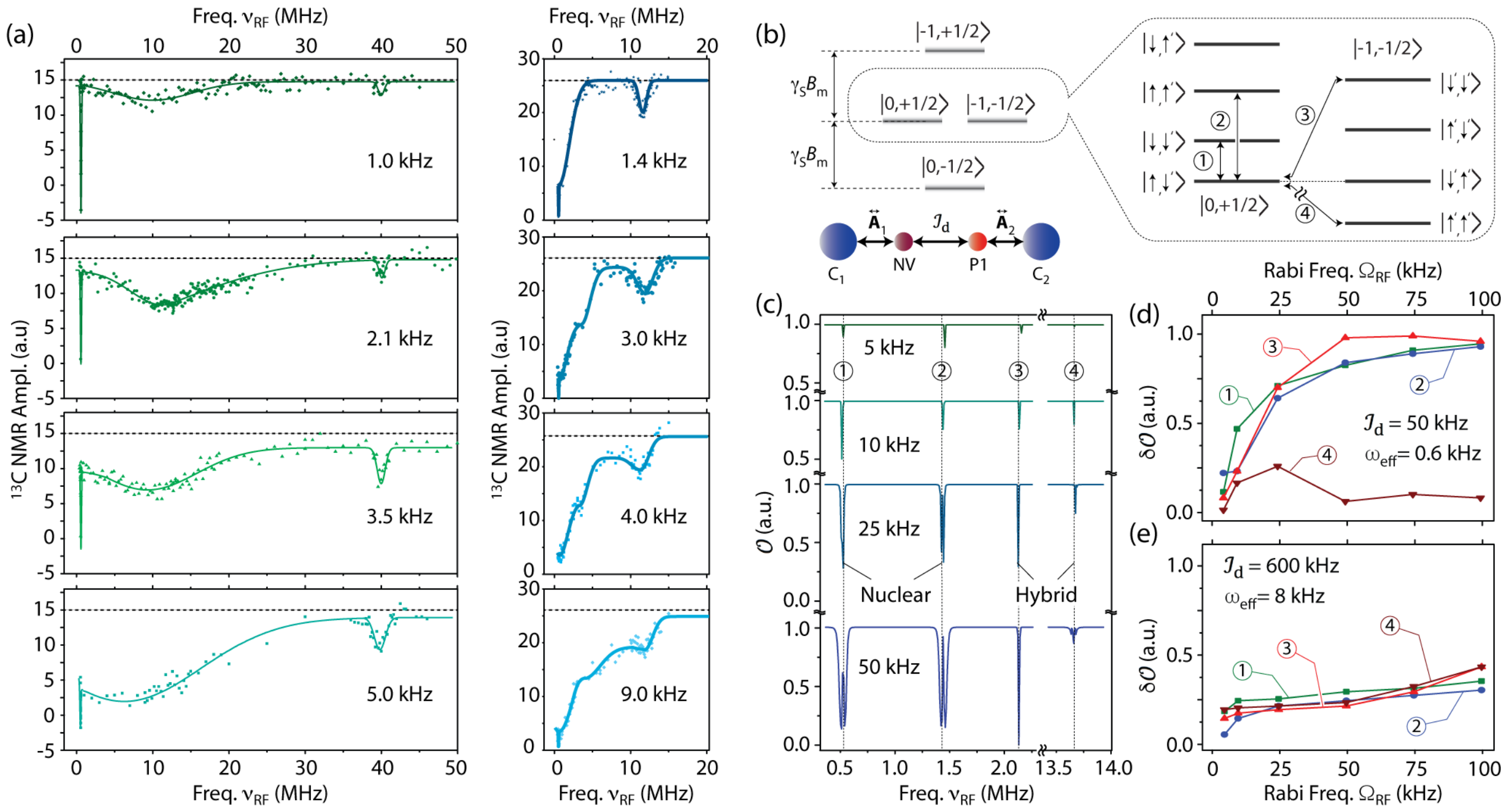

**Figure 3 | Dependence with RF power.** (a) $^{13}$C NMR signal amplitude as a function of the excitation frequency $\nu_{\mathrm{RF}}$ using the DNP protocols in Figs. 1a and 1d (respectively, left and right panels) for various RF amplitudes (bottom right in each panel). Horizontal dashed lines indicate the $^{13}$C NMR amplitude in the absence of RF excitation and solid traces are guides to the eye. (b) Schematic energy diagram for the electron-nuclear spin chain in the cartoon assuming the matching field $B = B_{\mathrm{m}}^{\alpha}$. States are denoted using projection numbers for the electronic spin and up/down arrows for nuclear spins with primes indicating a dominating hyperfine field. Numbers illustrate some nuclear and electron-nuclear spin transitions; energy separations are not to scale. (c) Spectral overlap $\mathcal{O}$ as a function of $\nu_{\mathrm{RF}}$ for different Rabi amplitudes $\Omega_{\mathrm{RF}}$ in the case of a spin chain with couplings $\mathcal{J}_{\mathrm{d}}$ = 30 kHz, $A_1^{xz}$ = 9 MHz, $A_2^{xz}$ = 2.5 MHz, and $A_j^{zz} = A_j^{zx}$ for $j = 1, 2$. (d) Spectral overlap change $\delta\mathcal{O}$ at select frequencies (bottom right) as a function of $\Omega_{\mathrm{RF}}$ for the spin chain in (c). (e) Same as in (d) but for a spin chain with couplings $\mathcal{J}_{\mathrm{d}}$ = 800 kHz, $A_1^{xz}$ = 9 MHz, $A_2^{xz}$ = 2.5 MHz, and $A_j^{zz} = A_j^{zx}$ for $j = 1, 2$.

While the model spin chain above fails to produce nuclear polarization transport away from the matching field range, we hypothesize that other, larger spin clusters featuring source and target $^{13}$C–NV dimers can still maintain transport through higher-order channels, though a formal description becomes increasingly complex[21,32]. One interesting example is the 5-spin chain $^{13}$C$_1$–NV$_1$–P1–NV$_2$–$^{13}$C$_2$ whose states $|\downarrow, 0, +1/2, -1, \uparrow\rangle$ and $|\uparrow, -1, +1/2, 0, \downarrow\rangle$ become degenerate when the inter-electronic dipolar coupling and hyperfine energies are suitably matched. Conversion of one into the other occurs via the virtual intermediate state $|\uparrow, -1, -1/2, -1, \uparrow\rangle$ at a rate of order $J'_{\mathrm{eff}} \sim \sin^2\left(\frac{\theta}{2}\right) \mathcal{J}_{\mathrm{d}}^2/(|\gamma_e|\, \delta B_{\mathrm{m}})$, where $\delta B_{\mathrm{m}}$ is the shift relative to the matching field. Note that because of the compensation between dipolar and hyperfine energies, large disparities between $\Delta_1$ and $\Delta_2$ (present only when at least one of the hyperfine couplings is large) cannot be easily accommodated by a reconfiguration of the electronic dipoles ($\mathcal{J}_d \lesssim 3$ MHz at the present paramagnetic center concentration). The result is that transport processes involving carbons strongly coupled to NVs get suppressed, in qualitative agreement with our observations. At the same time, polarization exchange remains efficient for moderately coupled nuclei: For example, for $\Delta_1 \sim \Delta_2 \sim 1$ MHz and $|\Delta_1 - \Delta_2| \sim 100$ kHz, we obtain $J'_{\mathrm{eff}} \sim 2$ kHz (we assume $\mathcal{J}_d \sim 1$ MHz and use $|\gamma_e|\, \delta B_{\mathrm{m}} \sim 120$ MHz, consistent with the conditions in Fig. 1e).

It is worth emphasizing that the increased degrees of freedom in the 5-spin set presented above are key to enabling inter-carbon spin transport, as a lengthy analysis of simpler chains shows; in particular, we find that no polarization exchange (other than the trivial case involving nuclei with identical hyperfine couplings) can take place away from the matching field if one or two electrons in the 5-spin chain are removed; the same is true if one of the NVs is replaced by a P1 (because the degeneracy between states involving different nuclear spin projections cannot be regained). Naturally, it is reasonable to expect transport contributions from other, more complex multi-spin arrays. Additional modeling and experiments (e.g., in the form of RF absorption spectra at fields farther removed from $B_\mathrm{m}$) will therefore be necessary to gain a fuller understanding.

In spite of the present limitations, we can tentatively interpret the markedly different frequency responses in Figs. 1b and 1e as the manifestation of two complementary spin transport regimes, one relying on field-enabled matching between NV and P1 resonances, the other emerging from P1-mediated interactions between NV-coupled carbons. A schematic is presented in Fig. 2e, where we generalize to more complex spin sets: $^{13}$C spins strongly coupled to NVs — otherwise thermalizing with the rest through the help of P1-based networks — become localized when the magnetic field departs sufficiently from $B_\mathrm{m}$. In this regime, dipolar P1-mediated interactions between NVs can help transport the polarization induced by chirped MW pulses in the (more-weakly-coupled) carbons in their vicinity. In particular, we hypothesize this latter mechanism underlies the disappearance or reduction of all dips above ~1 MHz in the RF absorption spectrum at 47.1 mT (Figs. 1e and 1f). Note that although chains involving only P1s — i.e., with no NVs — remain efficient spin exchange routes away from $B_\mathrm{m}$[13], such transport channels are not observable here because MW pulses selectively seed polarization in nuclei coupled to NVs, not P1s (i.e., an all-P1 chain can impact the NMR signal only in the less-likely scenario where the seed carbon is simultaneously coupled to an NV and a P1).

### D. Understanding the impact of RF on multi-spin electron/nuclear networks

Additional information on the dynamics at play can be obtained through the experiments in Fig. 3, where we measure the DNP response under the protocols of Figs. 1a and 1d using RF excitation of variable power. Besides the anticipated gradual growth of the absorption dips, we observe an overall spectral broadening, greatly exceeding that expected from increased RF power alone. This behavior is clearest in the range 5–15 MHz and near 40 MHz (Fig. 3a), where all absorption dips grow to encompass several MHz even when the RF Rabi field $\Omega_\mathrm{RF}$ never exceeds 10 kHz.

To interpret these observations, we resort one more time to the electron–nuclear spin chain in Fig. 2a and model the system dynamics in the presence of a driving RF field with no approximations[21,33]. Since optical initialization of the NV into $|0\rangle$ imposes a time dependence on the mean magnetization $\langle I_j^z\rangle$, $j = 1, 2$ of either nuclear spin in the chain[21], we gauge the impact of the drive at frequency $\nu_\mathrm{RF}$ and amplitude $\Omega_\mathrm{RF}$ via the overlap function $\mathcal{O}(\nu_\mathrm{RF}, \Omega_\mathrm{RF}) = \zeta \left| \int d\omega \, \langle I_1^z\rangle_\omega \langle I_2^z\rangle_\omega^* \right|$, where $\langle I_j^z\rangle_\omega = \int dt \, e^{i\omega t} \langle I_j^z\rangle(t, \nu_\mathrm{RF}, \Omega_\mathrm{RF})$ is the Fourier transform of the magnetization in carbon $j = 1, 2$, and $\zeta$ is a normalization constant calculated as the inverse of the spectral overlap $\left| \int d\omega \, \langle I_1^z\rangle_\omega \langle I_2^z\rangle_\omega^* \right|_0$, where the subscript denotes the absence of a drive (i.e., $\Omega_\mathrm{RF} = 0$). Maximum by default, $\mathcal{O}(\nu_\mathrm{RF}, \Omega_\mathrm{RF})$ decreases when $\nu_\mathrm{RF}$ is made resonant with one of the possible nuclear/electron spin transitions in the chain (see schematic energy diagram in Fig. 3b), thus allowing one to quantify the RF-induced disruption of transport through the appearance of 'dips' at select frequencies[21].

For illustration purposes, Fig. 3c shows the calculated response of a 4-spin chain with inter-electronic coupling $\mathcal{J}_\mathrm{d} = 30$ kHz assuming one of the transport-enabling conditions, $B = B_\mathrm{m}^{(\alpha)}$. RF-absorption at select frequencies perturbs inter-nuclear transport hence leading to a reduction of the spectral overlap $\mathcal{O}(\nu_\mathrm{RF}, \Omega_\mathrm{RF})$. A detailed inspection shows that some of these resonances can be associated to 'zero-quantum' (i.e., intra-band) transition frequencies in the electron bath. Normally forbidden, these transitions are activated here due to the hybrid, nuclear–electron spin nature of the chain (e.g., transitions (3) and (4) in Fig. 3b, see also Ref. 21). The separation between consecutive dips is determined by the inter-electron and hyperfine couplings, thus leading to complex spectral responses spanning several MHz.

Fig. 3d shows the calculated spectral overlap change $\delta\mathcal{O} \equiv \mathcal{O}(\Omega_\mathrm{RF}, \nu_\mathrm{RF}) - 1$ as a function of $\Omega_\mathrm{RF}$ at select excitation frequencies $\nu_\mathrm{RF}$: Interestingly, we find that all dips — both nuclear and hybrid — grow at comparable rates, a counter-intuitive response given the presumably hindered nature of the zero-quantum transitions[21]. On the other hand, the transport of nuclear spin polarization — faster for chains featuring greater $\mathcal{J}_\mathrm{d}$ — is more difficult to disrupt if $\Omega_\mathrm{RF} \lesssim J_\mathrm{eff}, J'_\mathrm{eff}$, thus leading to slower growth rates for more strongly coupled chains (Fig. 3e). Correspondingly, the response expected for spins in a crystal (vastly more complex than our toy model) is one where RF excitation of increasing amplitude gradually induces new dips through the perturbation of faster polarization transport channels. The result is a progressively broader-looking absorption spectrum, in qualitative agreement with our observations. Note that this picture also applies to the case where chirped MW excitation is simultaneously present (right panels in Fig. 3a), because the time interval (~2 ms) separating consecutive sweeps is typically longer than the inverse effective coupling, $(J'_\mathrm{eff})^{-1}$, thus ensuring the MW-induced disruption on polarization transport is minor.

## III. CONCLUSIONS

In summary, by monitoring changes in the DNP signal of $^{13}$C spins in diamond in the presence of an RF drive we show that hyperfine-coupled nuclei are central to the transport of spin polarization in the crystal. Further, different transport channels (involving nuclei featuring stronger or weaker hyperfine interactions) activate or not depending on the applied magnetic field. We conclude from this finding that the widespread notion of a spin-diffusion barrier intrinsic to the system under investigation is inaccurate, namely, strongly-hyperfine-coupled nuclei localize or delocalize depending on the 'connectivity' of interacting paramagnetic centers — itself a function of the defect concentration — here effectively controlled via the applied magnetic field.

Upon varying the amplitude of the drive, we witness gradual changes in the RF absorption spectrum — crudely manifesting as an overall broadening — which we analyze by considering the impact of continuous excitation on the dynamics of electron/nuclear spin chains. We find the RF drive disrupts nuclear spin transport through the activation of single- and many-spin transitions, the latter class involving both electron and nuclear spin flips. Our calculations show that systems featuring stronger inter-electronic couplings are less sensitive to RF excitation, indicating that the observed spectral changes stem from an inhomogeneous response where various spin sets — initially unaffected by weaker drives — gradually stop transporting nuclear polarization to the bulk as the RF amplitude grows. This view is consistent with the intuitive idea of multiple transport channels simultaneously coexisting in a disordered system.

Despite its present limitations, our model suggests we should view these many-spin sets as a single whole, where nominally forbidden 'hybrid' excitations applied locally propagate spectrally to impact groups of spins not directly addressed. Therefore, besides the fundamental aspects, an intriguing practical question is whether, even in the absence of optical pumping, Overhauser- or Solid-Effect-like DNP — normally relying on strong MW excitation — can be attained more simply via low-frequency (i.e., RF) manipulation of the electron spins using nuclei as local handles. More generally, these results could prove useful in quantum applications relying on spin platforms, for example, to transport information between remote nuclear qubits or to develop enhanced nanoscale sensing protocols.

## IV. ACKNOWLEDGEMENTS

D.P., J.H., and C.A.M. acknowledge support from the National Science Foundation through grant NSF-1903839, and from Research Corporation for Science Advancement through a FRED Award; they also acknowledge access to the facilities and research infrastructure of the NSF CREST IDEALS, grant number NSF-HRD-1547830. J.H. acknowledges support from CREST-PRF NSF-HRD 1827037. P.R.Z. and R.H.A. acknowledge financial support from CONICET (PIP-111122013010074 6CO), SeCyT-UNC (33620180100154CB) and ANPCYT (PICT-2014-1295).

**Supporting Information**

# *Magnetic-field-induced delocalization in hybrid electron-nuclear spin ensembles*

**Daniela Pagliero[1,*], Pablo R. Zangara[6,7,*], Jacob Henshaw[1], Ashok Ajoy[3], Rodolfo H. Acosta[6,7], Neil Manson[8], Jeffrey A. Reimer[4,5], Alexander Pines[3,5], Carlos A. Meriles[1,2]**

*[1] Department. of Physics, CUNY-City College of New York, New York, NY 10031, USA. [2] CUNY-Graduate Center, New York, NY 10016, USA. [3] Department of Chemistry, University of California at Berkeley, Berkeley, California 94720, USA. [4] Department of Chemical and Biomolecular Engineering, University of California at Berkeley, Berkeley, California 94720, USA. [5] Materials Science Division Lawrence Berkeley National Laboratory, Berkeley, California 94720, USA. [6] Facultad de Matemática, Astronomía, Física y Computación, Universidad Nacional de Córdoba, Ciudad Universitaria, CP:X5000HUA Córdoba, Argentina. [7] Instituto de Física Enrique Gaviola (IFEG), CONICET, Medina Allende s/n, X5000HUA, Córdoba, Argentina. [8] Laser Physics Centre, Research School of Physics and Engineering, Australian National University, Canberra, A.C.T. 2601, Australia.*

[*]Equally contributing authors

## I. The four-spin model

A simple spin system to study the main features of the magnetic-field controlled spin transport is presented in Fig. **S1**. Two $^{13}$C nuclear spins are hyperfine-coupled to two paramagnetic impurities, one of them a NV center and the other a P1 center. These two electronic spins, in turn, interact by means of a dipolar coupling. Given the typically large mismatch between the resonance frequencies of hyperfine-coupled and bulk carbons — and thus the corresponding quenching of inter-nuclear flip-flop processes — the 4-spin model above provides a rationale for understanding the dynamics of nuclear polarization in a real crystal. Following the arguments in Ref. [1], these mechanisms lead to an effective, purely nuclear, description of spin diffusion among a large number of $^{13}$Cs spins.

The Hamiltonian for the four-spin system is given by

$$H_{\mathrm{T}} = -\omega_{\mathrm{I}} I_1^z - \omega_{\mathrm{I}} I_2^z + \omega_e S^z + \omega_e S'^z + D(S^z)^2 + A_1^{zz} S^z I_1^z + A_1^{zx} S^z I_1^x + A_2^{zz} S'^z I_2^z + A_2^{zx} S'^z I_2^x + \frac{\mathcal{J}_{\mathrm{d}}}{2}\left(S^+ S'^+ + S^- S'^-\right). \tag{A.1}$$

Here, $\boldsymbol{I}_j$ $(j = 1, 2)$ stands for the nuclear $^{13}$C spin operator, $\boldsymbol{S}$ is the NV electronic spin operator ($S = 1$), $\boldsymbol{S}'$ is the P1 electronic spin operator ($S' = 1/2$), $\omega_{\mathrm{I}} = \gamma_{\mathrm{I}} B$, $\omega_e = |\gamma_e| B$, and $D$ corresponds to the NV zero-field splitting. Coefficients $A_{1(2)}^{zz}$ and $A_{1(2)}^{zx}$ respectively denote the hyperfine tensor components coupling the left (right) $^{13}$C and the NV (P1), and $\mathcal{J}_{\mathrm{d}}$ stands for the dipolar coupling strength between the NV and the P1 centers. In Eq. (A.1) both hyperfine interactions have already been secularized.

We assume the magnetic field is aligned with the NV axis and restrict our analysis to the vicinity of $B = 51$ mT, where the spin states $|0\uparrow\rangle$ for the NV-P1 pair is almost degenerate with $|-1\downarrow\rangle$. Such a degeneracy condition justifies the fact that we are only retaining the double-quantum terms in the NV-P1 dipolar interaction. Furthermore, we focus the analysis of the nuclear spin dynamics in the subspaces spanned by these two electronic states, since all other electronic configurations remain energetically inaccessible. Table **S1** shows the matrix representation of the Hamiltonian in Eq. (A.1) in such a subspace. The notation for the complete (electronic and nuclear) states is given by $\left|m_I^{(1)}, m_S, m_{S'}, m_I^{(2)}\right\rangle$.

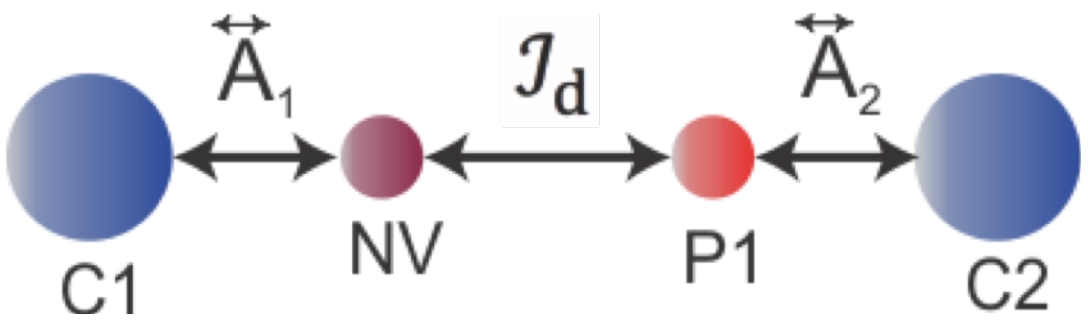

**Figure S1. The four-spin system.** Two [13]Cs interact with two dipolarly coupled electrons, one NV and one P1 center.

With the purpose of developing a model of effective $^{13}$C-$^{13}$C interactions, we perform a partial diagonalization of each $^{13}$C spin in a local basis given by the hyperfine interaction. More precisely, the quantization axis of each nuclear spin is now determined by the vectors

$$\vec{\mathcal{Z}}_1(m_S) = m_S A_1^{zx}\boldsymbol{i} + (m_S A_1^{zz} - \omega_\mathrm{I})\boldsymbol{k} \quad (A.2)$$

$$\vec{\mathcal{Z}}_2(m_{S'}) = m_{S'} A_2^{zx}\boldsymbol{i} + (m_{S'} A_2^{zz} - \omega_\mathrm{I})\boldsymbol{k} \quad (A.3)$$

for $^{13}$C spins C1 and C2, respectively. In the case of C1, notice that $\vec{\mathcal{Z}}_1$ introduces a non-trivial quantization axis only in the subspace corresponding to the electronic states $|-1\downarrow\rangle$, where $m_S = -1$. In the subspace with $|0\uparrow\rangle$ the quantization axis remains defined by the external magnetic field (no hyperfine interaction). This is not the case for $\vec{\mathcal{Z}}_2$ as it depends on $m_{S'} = \pm\frac{1}{2}$ (↑ or ↓). The norms of these vectors are related to the strength of the hyperfine interactions, and will be used in the following sections,

$$\Delta_1 = \left|\vec{\mathcal{Z}}_1(m_S = -1)\right| = \sqrt{(A_1^{zx})^2 + (A_1^{zz} + \omega_\mathrm{I})^2} \quad (A.4)$$

$$\Delta_2^{m_{S'}} = 2\left|\vec{\mathcal{Z}}_2(m_{S'})\right| = 2\sqrt{(m_{S'} A_2^{zx})^2 + (m_{S'} A_2^{zz} - \omega_\mathrm{I})^2}. \quad (A.5)$$

We can simplify these magnitudes by invoking the limit of strong hyperfine coupling $A_{1(2)}^{zz} \gg \omega_\mathrm{I}$. The dependence on both $\omega_\mathrm{I}$ and $m_S, m_{S'}$ can be dropped accordingly:

$$\Delta_1 \approx \sqrt{(A_1^{zx})^2 + (A_1^{zz})^2} \quad (A.6)$$

$$\Delta_2 \approx \Delta_2^{\uparrow} \approx \Delta_2^{\downarrow} \approx \sqrt{(A_2^{zx})^2 + (A_2^{zz})^2}. \quad (A.7)$$

The corresponding rotation angles required to transform into such an eigen-frame representation are respectively defined by

$$\tan(\theta_1) = A_1^{\mathrm{zx}}/(A_1^{zz} + \omega_\mathrm{I}) \quad (A.8)$$

$$\tan(\theta_2^{m_{S'}}) = m_{S'} A_2^{\mathrm{zx}}/(m_{S'} A_2^{zz} - \omega_\mathrm{I}). \quad (A.9)$$

Notice that in the case of $\theta_1$ the (electronic) spin quantum number $m_S = -1$ is omitted for simplicity.

The rotations of each nuclear quantization axis transform the original Hamiltonian $H_\mathrm{T}$ into a new one $\widetilde{H}_\mathrm{T}$, whose matrix representation is shown in Table **S2**. In this new basis, primed labels for the nuclear states indicate the change in the quantization axis when required.

## II. On the plausibility of single-electron-mediated interaction

In order to make our analysis as comprehensive as possible, we shall briefly digress and analyze here the plausibility of a simplified model where nuclear interactions are mediated by one, not two, electronic spins. We will show energy conservation makes, in general, this three-spin system incompatible with nuclear polarization exchange, thus making the four-spin chain considered above the simplest possible.

The simplified model of two nuclear spins with a single (P1) electron in the role of 'mediating particle' is described by the alternative Hamiltonian:

$$H_a = -\omega_I I_1^z - \omega_I I_2^z + \omega_e S'^z + A_1^{zz} S'^z I_1^z + A_1^{\perp}\left(S'^x I_1^x + S'^y I_1^y\right) + A_1^{zx} S'^z I_1^x + A_2^{zz} S'^z I_2^z +$$

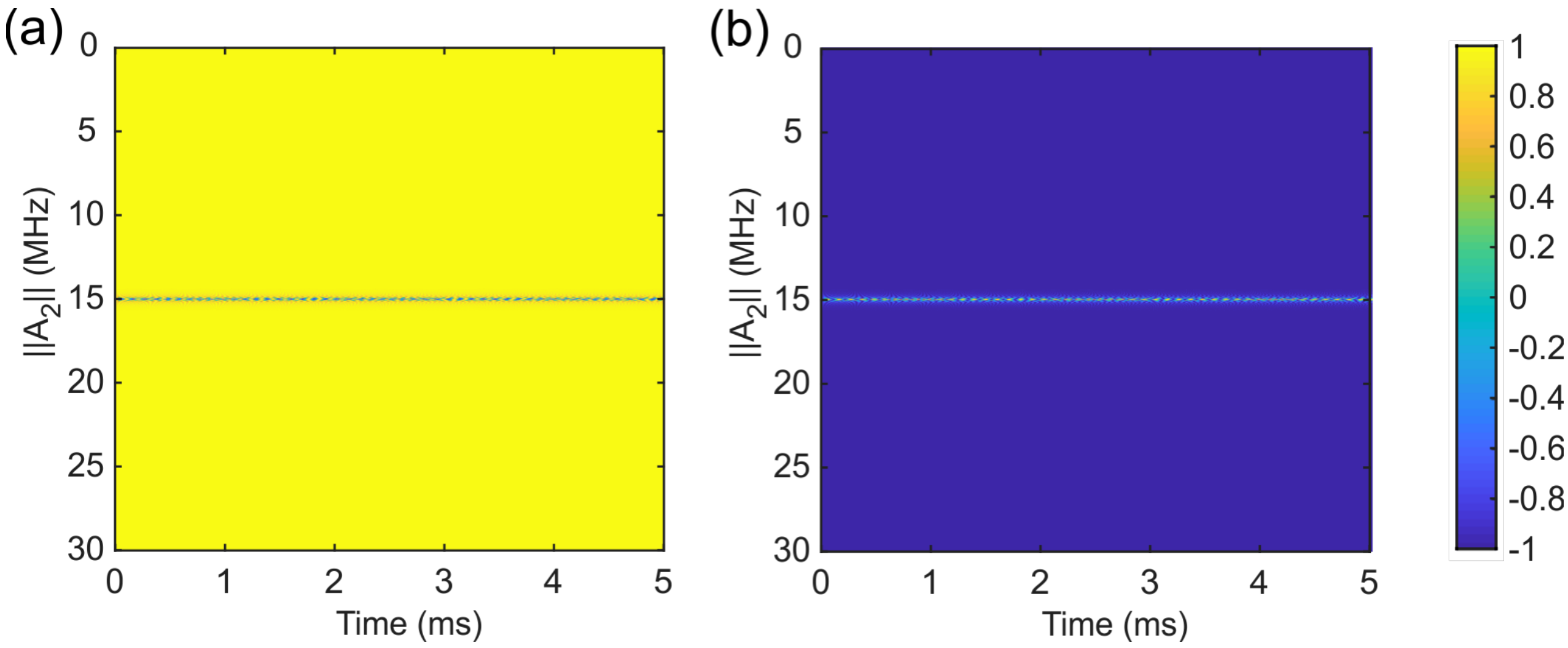

**Figure S2. Nuclear polarization dynamics in a $^{13}$C-P1-$^{13}$C chain.** (**a**) and (**b**) show the polarization of the first and second nuclear spins respectively, as a function of time and the strength of the second hyperfine coupling $\|A_2\|$. The first hyperfine coupling is maintained fixed, $\|A_1\| \equiv A_1^{zz} = A_1^{\perp} = A_1^{zx} = 15$ MHz. Dynamics corresponds to the Hamiltonian $\widetilde{H}_a$ as written in Table **S4**, the initial state is $|{\uparrow}'{\uparrow}{\downarrow}'\rangle$.

$$+A_2^{\perp}\left(S'^x I_2^x + S'^y I_2^y\right) + A_2^{zx} S'^z I_2^x. \tag{A.10}$$

Here we kept all possible terms in the two hyperfine interactions. The corresponding matrix representation of this Hamiltonian is shown in Table **S3** (states are labeled as $\left|m_I^{(1)}, m_{S'}, m_I^{(2)}\right\rangle$). Notice that $H_a$ naturally splits into two energetically separated blocks defined by the leading energy scale, $\omega_e$. This separation is what ultimately defines the 'secular' and 'non-secular' terms ($\propto A_1^{\perp}, A_2^{\perp}$). Following our prior strategy, we perform a partial diagonalization so the nuclear spin states are referred to their secular eigen-frames. The corresponding matrix representation, denoted $\widetilde{H}_a$, is shown in Table **S4**.

A quick inspection of the diagonal terms in $\widetilde{H}_a$ shows that the hyperfine and dipolar coupling parameters alone are insufficient to introduce a degeneracy between states $|{\uparrow}'{\uparrow}{\downarrow}'\rangle$ and $|{\downarrow}'{\uparrow}{\uparrow}'\rangle$ (or equivalently, between $|{\uparrow}'{\downarrow}{\downarrow}'\rangle$ and $|{\downarrow}'{\downarrow}{\uparrow}'\rangle$), except in the especial case where the hyperfine couplings are identical. Figure **S2** numerically confirms this conclusion via an explicit evaluation of the dynamics of nuclear spins in this system. We assume the initial state $|{\uparrow}'{\uparrow}{\downarrow}'\rangle$ and fix the first hyperfine coupling to be $\|A_1\| \equiv A_1^{zz} = A_1^{\perp} = A_1^{zx} =$ 15 MHz. Then we monitor the polarization of both nuclear spins as a function of time for different values of the hyperfine coupling $\|A_2\| \equiv A_2^{zz} = A_2^{\perp} = A_2^{zx}$; as anticipated, inter-carbon flip-flops occur only when $\|A_2\| \equiv \|A_1\|$. This restricted exchange is insufficient to rationalize polarization transport to bulk nuclei from strongly coupled carbons (as revealed by the experiments in Fig. 1 of the main text) and must, therefore, be ruled out.

## III. The effective $^{13}$C-$^{13}$C Hamiltonian

We resume here our discussion on the physical details of the four-spin model as presented in Fig. **S1**. It is clear from the matrix representation of Hamiltonian $\widetilde{H}_{\mathrm{T}}$ in Table **S2** that nuclear spin flip-flops are possible provided that the appropriate energy-matching condition is achieved. For instance, states $|{\uparrow}\ 0\ {\uparrow}{\downarrow}'\rangle$ and $|{\downarrow}\ {-1}\ {\downarrow}{\uparrow}'\rangle$ are degenerate if

$$-\frac{\omega_{\mathrm{I}}}{2} + \frac{\omega_e}{2} - \frac{\Delta_2^{\uparrow}}{4} = D - \frac{3\omega_e}{2} - \frac{\Delta_2^{\downarrow}}{4} + \frac{\Delta_1}{2}, \tag{A.11}$$

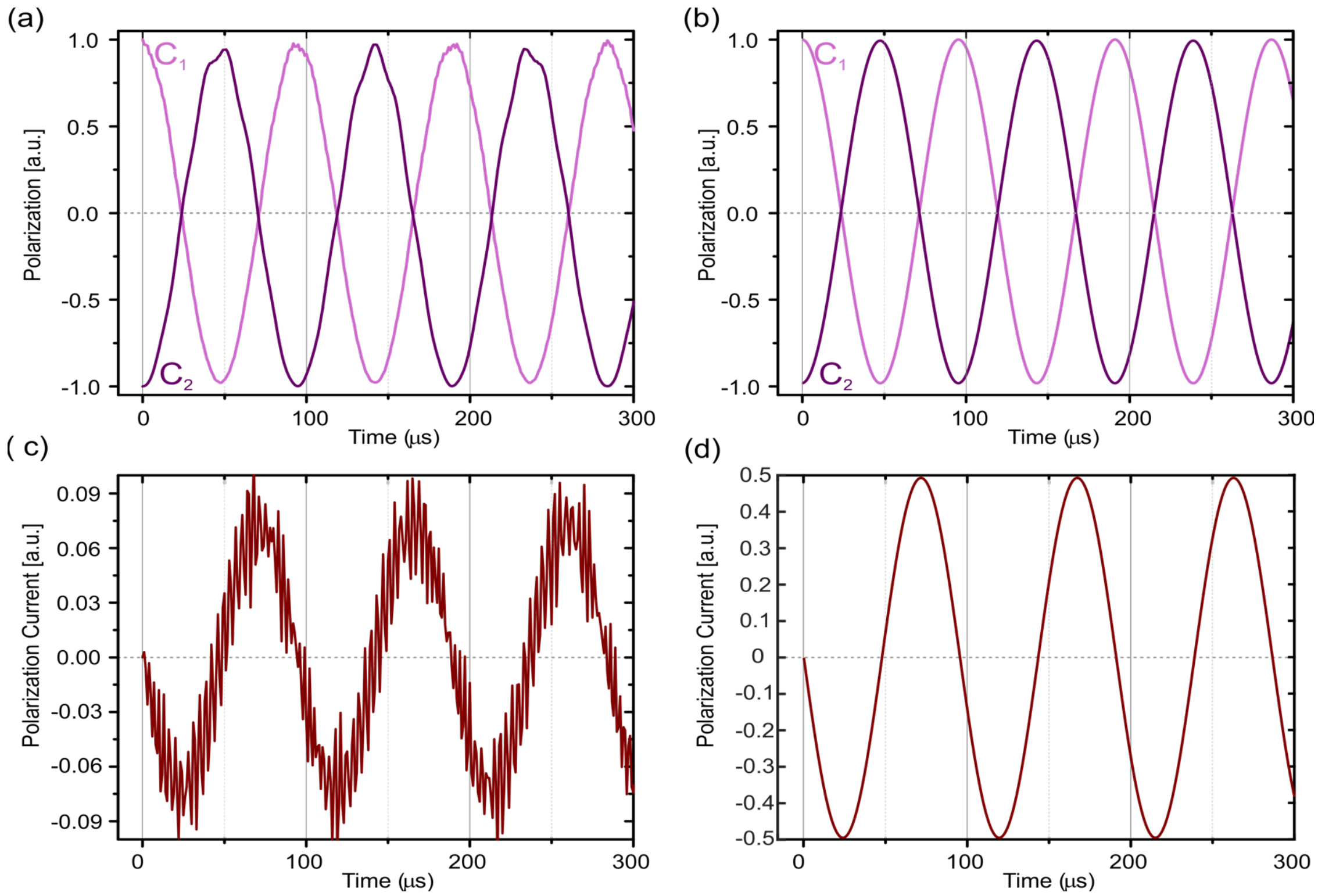

**Figure S3. Polarization and spin current dynamics in a $^{13}$C–NV–P1–$^{13}$C chain. (a)** Time dependence of polarization $P_1$ and $P_2$ at $^{13}$Cs C1 and C2 respectively, using the complete four-spin model evolving under Hamiltonian $\widetilde{H}_\mathrm{T}$ and initial state $|\uparrow 0 \uparrow\downarrow'\rangle$. (**b**) Same as in (**a**), but using the effective nuclear interaction given by Hamiltonian $H_\mathrm{eff}$, and initial state $|\uparrow\downarrow\rangle$. (**c**) Time dependence of the spin current for the dynamics induced by $\widetilde{H}_\mathrm{T}$ as in case (**a**). (**d**) Time dependence of the spin current for the effective interaction $H_\mathrm{eff}$. In all the cases, we consider $A_1^{zz} = A_1^{zx} = 9$ MHz, $A_2^{zz} = A_2^{zx} = 2.5$ MHz, $\mathcal{J}_\mathrm{d} = 800$ kHz, $B = B_m^{(\alpha)} = 51.3085$ mT.

which defines an equation for the 'matching' magnetic field $B = B_m^{(\alpha)}$. When this condition is met, the dynamics of the pair of nuclear spins is essentially a flip-flop $|\uparrow\downarrow'\rangle \leftrightarrow |\downarrow\uparrow'\rangle$. The time-scale for such process is given by the matrix element

$$\langle\downarrow -1 \downarrow\uparrow'|\widetilde{H}_\mathrm{T}|\uparrow 0 \uparrow\downarrow'\rangle = -\frac{\mathcal{J}_\mathrm{d}}{2}\sin(\theta_1/2)\sin[(\theta_2^{\downarrow} - \theta_2^{\uparrow})/2] \quad (\mathrm{A.12})$$

An equivalent situation can be found when the condition

$$\frac{\omega_\mathrm{I}}{2} + \frac{\omega_e}{2} + \frac{\Delta_2^{\uparrow}}{4} = D - \frac{3\omega_e}{2} + \frac{\Delta_2^{\downarrow}}{4} - \frac{\Delta_1}{2} \quad (\mathrm{A.13})$$

is satisfied. In this case, states $|\downarrow 0 \uparrow\uparrow'\rangle$ and $|\uparrow -1 \downarrow\downarrow'\rangle$ are degenerate, and the coupling matrix element is the same as in Eq. (A.12). This energy-matching condition corresponds to a different magnetic field $B = B_m^{(\beta)}$ as Eq. (A.11) is not equivalent to (A.13). At any of these 'matching' fields, an effective description can be proposed for the nuclear spins,

$$H_\mathrm{eff} = \delta_\mathrm{eff} I_1^z - \delta_\mathrm{eff} I_2^z + J_\mathrm{eff}(I_1^+ I_2^- + I_1^- I_2^+), \quad (\mathrm{A.14})$$

where

$$\delta_\mathrm{eff} = 2\left|B - B_m^{(\alpha,\beta)}\right|\gamma_e \quad (\mathrm{A.15})$$

$$J_{\mathrm{eff}} = \frac{\mathcal{J}_{\mathrm{d}}}{2}\sin(\theta_1/2)\sin[(\theta_2^{\downarrow} - \theta_2^{\uparrow})/2]. \tag{A.16}$$

Figure **S3**(**a-b**) compares the calculated [13]C polarization dynamics for both the complete $\widetilde{H}_{\mathrm{T}}$ and the effective $H_{\mathrm{eff}}$ Hamiltonians.

From Eq. (A.12) we conclude that the effective [13]C-[13]C coupling $J_{\mathrm{eff}}$ embodies a four-body transition matrix element. This is equivalent to the "hyperfine-dominated" effective coupling described in Ref. [1], where two P1 centers were used to mediate the [13]C spins. As before, we can use here the assumption of a 'hyperfine dominated' regime (i.e. the hyperfine energy is much larger than the nuclear Zeeman energy) to derive a more explicit form for $J_{\mathrm{eff}}$. In the limit $|\theta_2^{\downarrow} - \theta_2^{\uparrow}| \to 0$,

$$\sin\left[\frac{\theta_2^{\downarrow} - \theta_2^{\uparrow}}{2}\right] \approx \frac{1}{2}\tan\left(\theta_2^{\downarrow} - \theta_2^{\uparrow}\right) = \left(\frac{1}{2}\right)\frac{\tan\theta_2^{\downarrow} - \tan\theta_2^{\uparrow}}{1 + \tan\theta_2^{\downarrow}\tan\theta_2^{\uparrow}}$$

$$= \left(\frac{1}{2}\right)\frac{\left[\frac{-\frac{1}{2}A_2^{zx}}{\left(-\frac{1}{2}A_2^{zz} - \omega_I\right)}\right] - \left[\frac{+\frac{1}{2}A_2^{zx}}{\left(+\frac{1}{2}A_2^{zz} - \omega_I\right)}\right]}{1 + \frac{\frac{1}{4}(A_2^{zx})^2}{\left[\left(\frac{1}{2}A_2^{zz}\right)^2 - {\omega_I}^2\right]}} \tag{A.17}$$

Then,

$$\sin\left[\frac{\theta_2^{\downarrow} - \theta_2^{\uparrow}}{2}\right] \approx \frac{-2\omega_I A_2^{zx}}{[(A_2^{zz})^2 - (2\omega_I)^2] + (A_2^{zx})^2}. \tag{A.18}$$

In this limit, we safely drop the dependence on $\omega_I$ in the denominator, and rewrite the effective flip-flop matrix element as

$$J_{\mathrm{eff}} \approx -\mathcal{J}_{\mathrm{d}}\sin\left(\frac{\theta_1}{2}\right)\frac{\omega_I A_2^{zx}}{(A_2^{zz})^2 + (A_2^{zx})^2}, \tag{A.19}$$

which is analogous to the matrix element for the case of a pair of P1 centers in the 'mediating' role,

$$J_{\mathrm{eff}} \approx -\mathcal{J}_{\mathrm{d}}\left(\frac{\omega_I A_1^{zx}}{(A_1^{zz})^2 + (A_1^{zx})^2}\right)\left(\frac{\omega_I A_2^{zx}}{(A_2^{zz})^2 + (A_2^{zx})^2}\right), \tag{A.20}$$

as derived in Ref. [1]. We warn, however, that the P1-P1 mechanism of electron-mediated nuclear interaction is not dependent on the magnetic field (more precisely, it does not require the field-matching condition), so it is insufficient for rationalizing our experimental observations.

## IV. The spin current and the delocalization diagram

A complementary analysis of the polarization dynamics can be done in terms of the polarization current [2]

$$\mathcal{K} = (1/2i)(I_1^- I_2^+ - I_1^+ I_2^-), \tag{A.21}$$

which provides an observable to quantify the flow of polarization between the two [13]C spins. For example, we show in Fig. **S3** the time-dependence of the polarization at each [13]C, $P_j(t) = 2\mathrm{tr}(I_j^z\rho(t)), j = 1,2$, along with $\mathcal{K}$ for a given set of couplings and satisfying the matching condition $B = B_m^{(\alpha)}$.

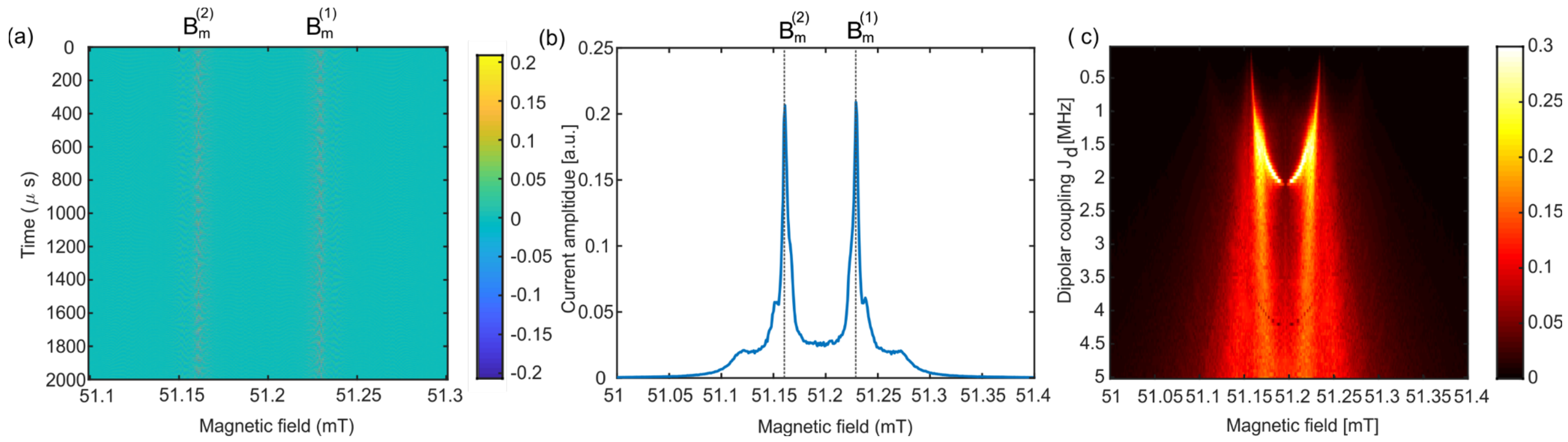

**Figure S4**. **Dynamics of spin current $\mathcal{K}$.** (**a**) Time evolution of $\mathcal{K}$ as a function of the magnetic field. (**b**) Amplitude of the oscillations in $\mathcal{K}$ (i.e., maximum along the time-axis in (**a**)) as a function of the magnetic field. (**c**) Same as in (**b**), but for a variable dipolar coupling strength $\mathcal{J}_\mathrm{d}$. In all the cases, we consider $A_1^{zz} = A_1^{zx} = 2.9$ MHz, $A_2^{zz} = A_2^{zx} = 3.2$ MHz; the initial state is given in Eq. (A.22). In (**a-b**), $\mathcal{J}_\mathrm{d} = 800$ kHz.

Our model being strictly finite, both the spin polarization and the current keep oscillating as the two nuclear spins undergo flip-flops. A more realistic description would encompass a large number of nuclear spins effectively interacting by means of an appropriate extension of Hamiltonian $H_{\mathrm{eff}}$ in Eq. (A.14). In such a case, the polarization would jump from the second $^{13}$C to a nearby $^{13}$C (also coupled by mediating NV-P1 or P1-P1 pairs), ultimately diffusing away. This physical picture of spin-diffusion from strongly hyperfine-shifted $^{13}$Cs to gradually more weakly hyperfine-shifted $^{13}$Cs (and then finally to bulk $^{13}$Cs) has been extensively discussed Ref. [1].

In Fig. **S4**(**a**) we evaluate $\mathcal{K}$ as a function of time and the magnetic field $B$ for a given choice of hyperfine couplings and $\mathcal{J}_\mathrm{d}$. In order to capture the dynamical trends at both 'matching' fields $B_m^{(\alpha)}$ and $B_m^{(\beta)}$, we consider the initial state

$$\rho_0 = \frac{1}{2}|\uparrow 0 \uparrow\downarrow'\rangle\langle\uparrow 0 \uparrow\downarrow'| + \frac{1}{2}|\downarrow 0 \uparrow\uparrow'\rangle\langle\downarrow 0 \uparrow\uparrow'|. \tag{A.22}$$

Figure **S4**(**b**) shows a cross-section of the amplitude of the oscillations in $\mathcal{K}$ as a function of $B$. Notice

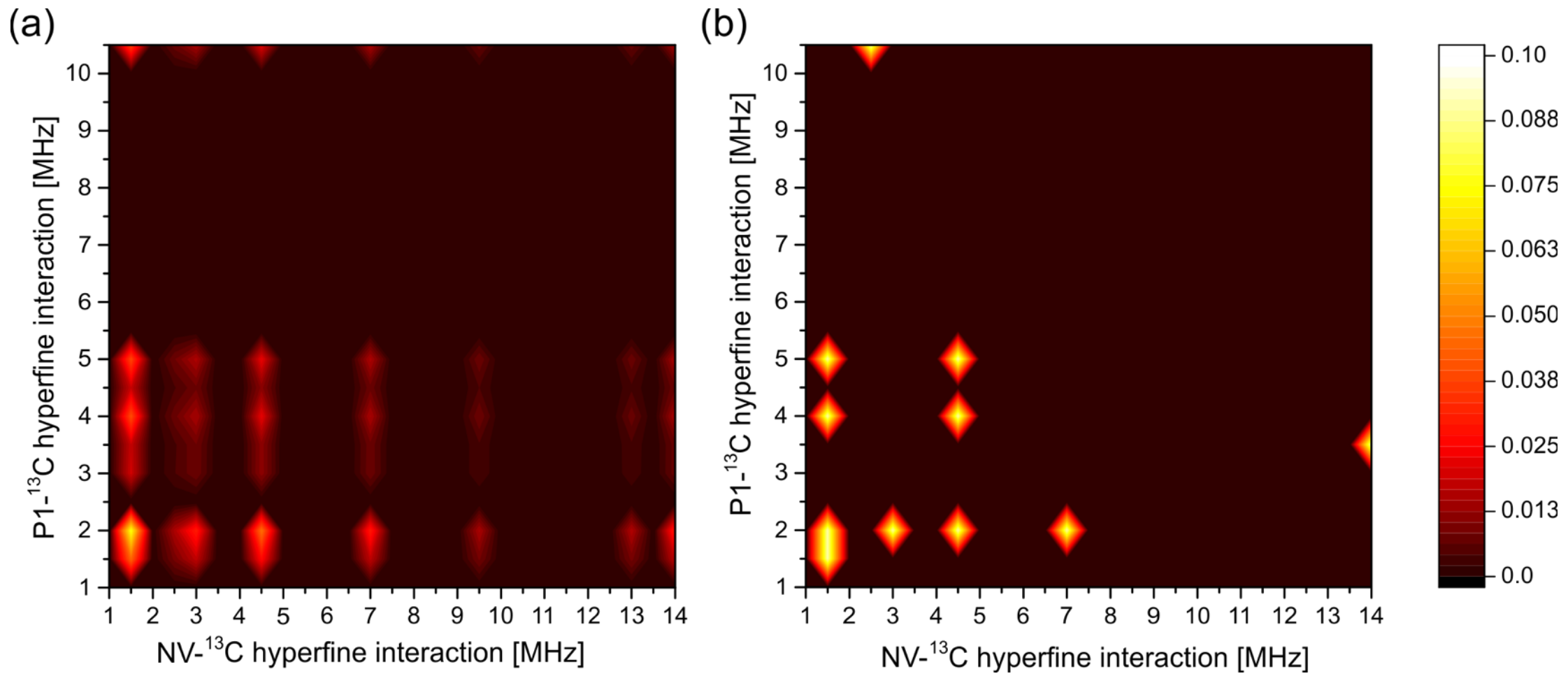

**Figure S5**. **Distributions of hyperfine couplings.** (**a**) Bivariate distribution of NV-$^{13}$C and P1-$^{13}$C hyperfine couplings as reported in Refs. [3-4] and [5], respectively. (**b**) Bivariate 'toy' distribution employed to mimic (**a**) and compute the averaged case shown in Fig. 2(d) of the main text.

the two dominant peaks corresponding to $B_m^{(\alpha)}$ and $B_m^{(\beta)}$, whose widths are associated to higher order processes. Figure **S4**(**c**) shows the same quantity, the amplitude of $\mathcal{K}$ as a function of $B$, but here calculated for different $\mathcal{J}_{\rm d}$ (Figure 2(c) in the main text shows the same simulation for different sets of hyperfine couplings).

The amplitude of $\mathcal{K}$ is used here as a quantitative indicator for localization/delocalization of nuclear spin polarization. As the amplitude of $\mathcal{K}$ increases, the polarization is more efficiently transferred and it can therefore diffuse. This leads us to build a dynamical phase-diagram as shown in Fig. 2(d) of the main text, where we average over many different possible choices of hyperfine couplings (configurations in our 4-spin system). Since the possible NV-$^{13}$C and P1-$^{13}$C hyperfine couplings are well known [3–5], we performed a sampling of the most relevant combinations of couplings from 1 to ~14 MHz. In particular, Fig. **S5**(**a**) shows the actual bivariate distribution of possible hyperfine configurations and Fig. **S5**(**b**) shows the cases we evaluated and accounted for to mimic such distribution.

## V. The 4-spin system in the presence of RF driving

In order to analyze the spectral broadening observed in the DNP signal under RF excitation, we study the dynamics of polarization transfer in our 4-spin model adding the time-dependent perturbation

$$V(t) = \Omega_{RF}\cos(\nu_{RF}t)(I_1^x + I_2^x), \tag{A.23}$$

where $\Omega_{RF}$ stands for RF amplitude and $\nu_{RF}$ denotes its frequency. The system's dynamical response is computed without any explicit approximation using the QuTiP [6] time-dependent solver.

Fig. **S6**(**a-b**) shows the time dependence of the polarization of each carbon $P_j(t)$ ($j = 1,2$), as a function of the irradiation frequency $\nu_{RF}$. Here, we assume resonant polarization transfer (i.e., $B = B_m^{(\alpha)}$) and a weak effective interaction $J_{\rm eff}$~1 kHz ($A_1^{zz} = A_1^{zx} = 9$ MHz, $A_2^{zz} = A_2^{zx} = 2.5$ MHz, $\mathcal{J}_{\rm d} = 30$ kHz). The amplitude $\Omega_{\rm RF}$ is kept constant at 25 kHz.

Four possible transitions can be observed as distortions in the synchronized flip-flop dynamics shown in Fig. **S6**(**a-b**), each of them distinguished with a label (1,2,3,4). Using the Hamiltonian representation in Table **S2**, these transitions can be identified according to Table **S5**.

In order to assess the effect of RF excitation, we compute the Fourier transform of $P_j(t)$ for a given set of conditions ($\Omega_{RF}, \nu_{RF}$). Each normalized Fourier spectrum $P_j(\omega)$ can be then compared to quantify how the two $^{13}$C spins decouple in the presence of a time-dependent perturbation. Figs. **S6**(**c-d**) show the unperturbed spectra $P_1^{(0)}(\omega)$ and $P_2^{(0)}(\omega)$ for C1 and C2, respectively, obtained in the absence of RF irradiation. Figs. **S6**(**e-f**) show the perturbed spectra $P_1(\omega;\nu_{RF})$ and $P_2(\omega;\nu_{RF})$ for C1 and C2 respectively, obtained in the presence of RF excitation at a frequency $\nu_{RF}$ corresponding to transition #3 and a given (fixed) RF amplitude $\Omega_{\rm RF}$. This approach provides a systematic way to evaluate the decoupling of the two $^{13}$C spins, by computing the overlap between the Fourier spectra $P_j(\omega)$. Indeed, in cases (**c-d**) the spectral overlap is ~1 (as in the case of non-resonant RF excitation). Quite on the contrary, the overlap is less than 1 in cases (**e-f**), which means that the two $^{13}$C nuclei are, to some extent, decoupled. This magnitude is ultimately associated to the 'dip' in the obtained signal, as shown in Fig. 3 of the main text.

| Label | States involved | Frequency |
|---|---|---|
| 1 | $\lvert\uparrow 0 \uparrow\downarrow'\rangle \leftrightarrow \lvert\downarrow 0 \uparrow\downarrow'\rangle$ | $\nu_{RF} \sim \omega_{\rm I} + \mathcal{O}({\mathcal{J}_{\rm d}}^2)$ |
| 2 | $\lvert\uparrow 0 \uparrow\downarrow'\rangle \leftrightarrow \lvert\uparrow 0 \uparrow\uparrow'\rangle$ | $\nu_{RF} \sim \Delta_2^{\uparrow}/2 + \mathcal{O}({\mathcal{J}_{\rm d}}^2)$ |
| 3 | $\lvert\uparrow 0 \uparrow\downarrow'\rangle \leftrightarrow \lvert\downarrow -1 \downarrow\downarrow'\rangle$ | $\nu_{RF} \sim \Delta_2^{\downarrow}/2 + \mathcal{O}({\mathcal{J}_{\rm d}}^2)$ |
| 4 | $\lvert\uparrow 0 \uparrow\downarrow'\rangle \leftrightarrow \lvert\uparrow -1 \downarrow\uparrow'\rangle$ | $\nu_{RF} \sim \Delta_1 + \mathcal{O}({\mathcal{J}_{\rm d}}^2)$ |

Table **S5**. Transition frequencies excited by the RF irradiation (see Fig. **S6**).

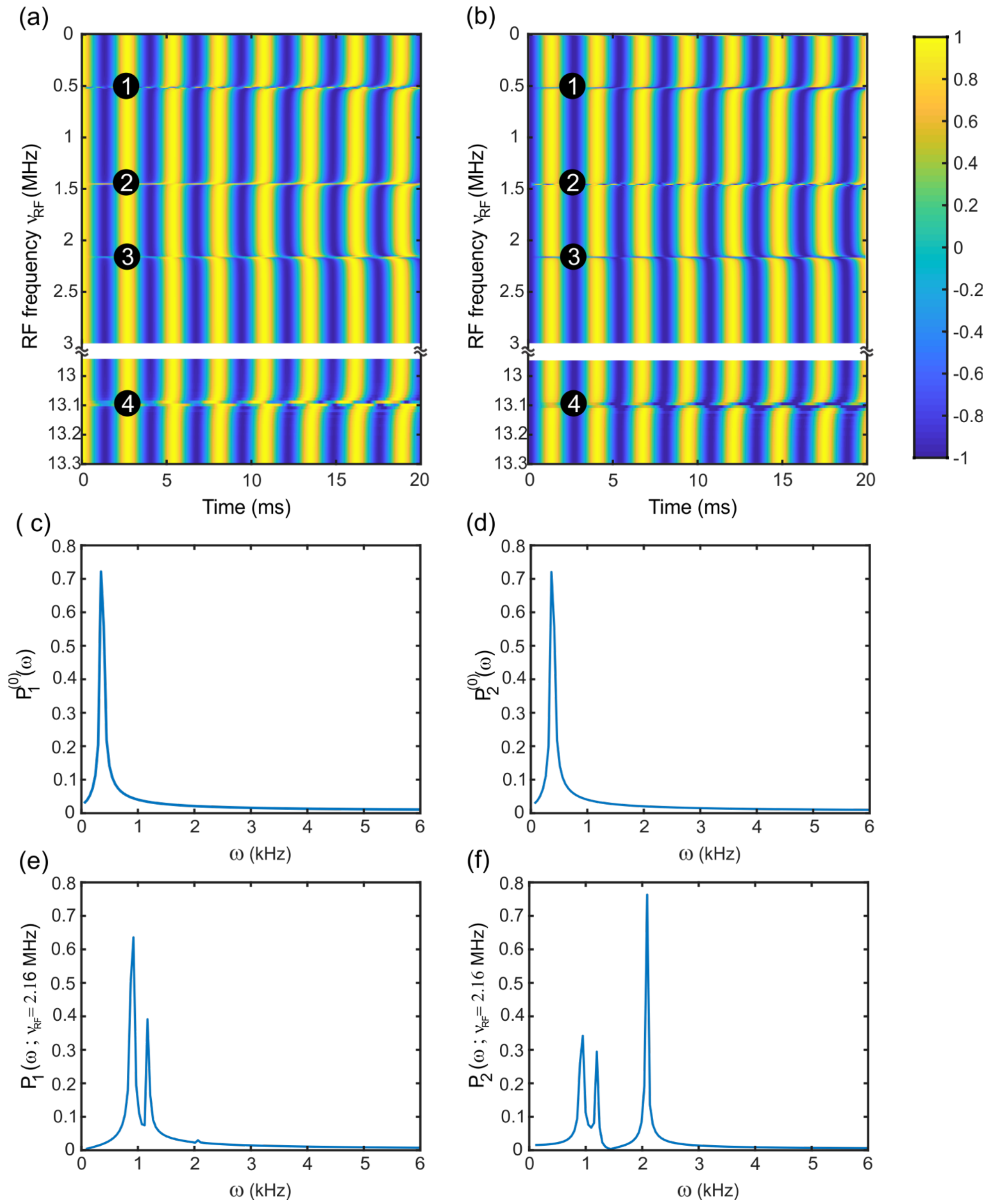

**Figure S6. Polarization dynamics in the presence of RF irradiation.** (**a-b**) Time response $P_j(t)$ as a function of the irradiation frequency $\nu_{RF}$ for C1 (**a**) and C2 (**b**). (**c-d**) Fourier spectra $P_1^{(0)}(\omega)$ and $P_2^{(0)}(\omega)$ for C1 and C2 respectively (no RF irradiation). (**e-f**) Fourier spectra $P_1(\omega;\nu_{RF})$ and $P_2(\omega;\nu_{RF})$ for C1 and C2 respectively, in the presence of RF irradiation with $\nu_{\mathrm{RF}} \approx 2.16$ MHz and $\Omega_{\mathrm{RF}} = 25$ kHz. In all cases, $A_1^{zz} = A_1^{zx} = 9$ MHz, $A_2^{zz} = A_2^{zx} = 2.5$ MHz, and $\mathcal{J}_{\mathrm{d}} = 30$ kHz.

| | $\|\uparrow 0 \uparrow\uparrow\rangle$ | $\|\uparrow 0 \uparrow\downarrow\rangle$ | $\|\downarrow 0 \uparrow\uparrow\rangle$ | $\|\downarrow 0 \uparrow\downarrow\rangle$ | $\|\uparrow -1 \downarrow\uparrow\rangle$ | $\|\uparrow -1 \downarrow\downarrow\rangle$ | $\|\downarrow -1 \downarrow\uparrow\rangle$ | $\|\downarrow -1 \downarrow\downarrow\rangle$ |
|---|---|---|---|---|---|---|---|---|
| $\langle\uparrow 0 \uparrow\uparrow\|$ | $-\omega_I + \frac{\omega_e}{2} + \frac{A_2^{zz}}{4}$ | $\frac{A_2^{zx}}{4}$ | 0 | 0 | $\mathcal{J}_d/2$ | 0 | 0 | 0 |
| $\langle\uparrow 0 \uparrow\downarrow\|$ | $\frac{A_2^{zx}}{4}$ | $\frac{\omega_e}{2} - \frac{A_2^{zz}}{4}$ | 0 | 0 | 0 | $\mathcal{J}_d/2$ | 0 | 0 |
| $\langle\downarrow 0 \uparrow\uparrow\|$ | 0 | 0 | $\frac{\omega_e}{2} + \frac{A_2^{zz}}{4}$ | $\frac{A_2^{zx}}{4}$ | 0 | 0 | $\mathcal{J}_d/2$ | 0 |
| $\langle\downarrow 0 \uparrow\downarrow\|$ | 0 | 0 | $\frac{A_2^{zx}}{4}$ | $\omega_I + \frac{\omega_e}{2} - \frac{A_2^{zz}}{4}$ | 0 | 0 | 0 | $\mathcal{J}_d/2$ |
| $\langle\uparrow -1 \downarrow\uparrow\|$ | $\mathcal{J}_d/2$ | 0 | 0 | 0 | $-\omega_I + D - \frac{3\omega_e}{2} - \frac{A_2^{zz}}{4} - \frac{A_1^{zz}}{2}$ | $-\frac{A_2^{zx}}{4}$ | $-\frac{A_1^{zx}}{2}$ | 0 |
| $\langle\uparrow -1 \downarrow\downarrow\|$ | 0 | $\mathcal{J}_d/2$ | 0 | 0 | $-\frac{A_2^{zx}}{4}$ | $D - \frac{3\omega_e}{2} + \frac{A_2^{zz}}{4} - \frac{A_1^{zz}}{2}$ | 0 | $-\frac{A_1^{zx}}{2}$ |
| $\langle\downarrow -1 \downarrow\uparrow\|$ | 0 | 0 | $\mathcal{J}_d/2$ | 0 | $-\frac{A_1^{zx}}{2}$ | 0 | $D - \frac{3\omega_e}{2} - \frac{A_2^{zz}}{4} + \frac{A_1^{zz}}{2}$ | $-\frac{A_2^{zx}}{4}$ |
| $\langle\downarrow -1 \downarrow\downarrow\|$ | 0 | 0 | 0 | $\mathcal{J}_d/2$ | 0 | $-\frac{A_1^{zx}}{2}$ | $-\frac{A_2^{zx}}{4}$ | $\omega_I + D - \frac{3\omega_e}{2} + \frac{A_2^{zz}}{4} + \frac{A_1^{zz}}{2}$ |

**Table S1**. **Matrix representation of Hamiltonian $\boldsymbol{H}_{\mathbf{T}}$**. The two subspaces corresponding to each NV-P1 electronic configuration have been shadowed with green ($|0\uparrow\rangle$) and grey ($|-1\downarrow\rangle$).

| | $\lvert\uparrow 0 \uparrow\uparrow'\rangle$ | $\lvert\uparrow 0 \uparrow\downarrow'\rangle$ | $\lvert\downarrow 0 \uparrow\uparrow'\rangle$ | $\lvert\downarrow 0 \uparrow\downarrow'\rangle$ | $\lvert\uparrow -1 \downarrow\uparrow'\rangle$ | $\lvert\uparrow -1 \downarrow\downarrow'\rangle$ | $\lvert\downarrow -1 \downarrow\uparrow'\rangle$ | $\lvert\downarrow -1 \downarrow\downarrow'\rangle$ |
|---|---|---|---|---|---|---|---|---|
| $\langle\uparrow 0 \uparrow\uparrow'\rvert$ | $-\frac{\omega_I}{2}+\frac{\omega_e}{2}+\frac{\Delta_2^\uparrow}{4}$ | 0 | 0 | 0 | $\frac{\mathcal{J}_d}{2}c_2c_1$ | $-\frac{\mathcal{J}_d}{2}s_2c_1$ | $-\frac{\mathcal{J}_d}{2}c_2s_1$ | $\frac{\mathcal{J}_d}{2}s_2s_1$ |
| $\langle\uparrow 0 \uparrow\downarrow'\rvert$ | 0 | $-\frac{\omega_I}{2}+\frac{\omega_e}{2}-\frac{\Delta_2^\uparrow}{4}$ | 0 | 0 | $\frac{\mathcal{J}_d}{2}s_2c_1$ | $\frac{\mathcal{J}_d}{2}c_2c_1$ | $-\frac{\mathcal{J}_d}{2}s_2s_1$ | $-\frac{\mathcal{J}_d}{2}c_2s_1$ |
| $\langle\downarrow 0 \uparrow\uparrow'\rvert$ | 0 | 0 | $\frac{\omega_I}{2}+\frac{\omega_e}{2}+\frac{\Delta_2^\uparrow}{4}$ | 0 | $\frac{\mathcal{J}_d}{2}c_2s_1$ | $-\frac{\mathcal{J}_d}{2}s_2s_1$ | $\frac{\mathcal{J}_d}{2}c_2c_1$ | $-\frac{\mathcal{J}_d}{2}s_2c_1$ |
| $\langle\downarrow 0 \uparrow\downarrow'\rvert$ | 0 | 0 | 0 | $\frac{\omega_I}{2}+\frac{\omega_e}{2}-\frac{\Delta_2^\uparrow}{4}$ | $\frac{\mathcal{J}_d}{2}s_2s_1$ | $\frac{\mathcal{J}_d}{2}c_2s_1$ | $\frac{\mathcal{J}_d}{2}s_2c_1$ | $\frac{\mathcal{J}_d}{2}c_2c_1$ |
| $\langle\uparrow -1 \downarrow\uparrow'\rvert$ | $\frac{\mathcal{J}_d}{2}c_2c_1$ | $\frac{\mathcal{J}_d}{2}s_2c_1$ | $\frac{\mathcal{J}_d}{2}c_2s_1$ | $\frac{\mathcal{J}_d}{2}s_2s_1$ | $D-\frac{3\omega_e}{2}-\frac{\Delta_2^\downarrow}{4}-\frac{\Delta_1}{2}$ | 0 | 0 | 0 |
| $\langle\uparrow -1 \downarrow\downarrow'\rvert$ | $-\frac{\mathcal{J}_d}{2}s_2c_1$ | $\frac{\mathcal{J}_d}{2}c_2c_1$ | $-\frac{\mathcal{J}_d}{2}s_2s_1$ | $\frac{\mathcal{J}_d}{2}c_2s_1$ | 0 | $D-\frac{3\omega_e}{2}+\frac{\Delta_2^\downarrow}{4}-\frac{\Delta_1}{2}$ | 0 | 0 |
| $\langle\downarrow -1 \downarrow\uparrow'\rvert$ | $-\frac{\mathcal{J}_d}{2}c_2s_1$ | $-\frac{\mathcal{J}_d}{2}s_2s_1$ | $\frac{\mathcal{J}_d}{2}c_2c_1$ | $\frac{\mathcal{J}_d}{2}s_2c_1$ | 0 | 0 | $D-\frac{3\omega_e}{2}-\frac{\Delta_2^\downarrow}{4}+\frac{\Delta_1}{2}$ | 0 |
| $\langle\downarrow -1 \downarrow\downarrow'\rvert$ | $\frac{\mathcal{J}_d}{2}s_2s_1$ | $-\frac{\mathcal{J}_d}{2}c_2s_1$ | $-\frac{\mathcal{J}_d}{2}s_2c_1$ | $\frac{\mathcal{J}_d}{2}c_2c_1$ | 0 | 0 | 0 | $D-\frac{3\omega_e}{2}+\frac{\Delta_2^\downarrow}{4}+\frac{\Delta_1}{2}$ |

**Table S2**. **Matrix representation for the Hamiltonian $\widetilde{\boldsymbol{H}}_{\mathbf{T}}$.** Here, $\Delta_1$ and $\Delta_2^{\uparrow(\downarrow)}$ are defined in Eqns. (A4-A5). In addition, $c_1 = \cos(\theta_1/2)$, $s_1 = \sin(\theta_1/2)$, $c_2 = \cos[(\theta_2^\downarrow - \theta_2^\uparrow)/2]$, and $s_2 = \sin[(\theta_2^\downarrow - \theta_2^\uparrow)/2]$, where $\theta_1$, $\theta_2^\downarrow$ and $\theta_2^\uparrow$ are defined in Eqns. (A.8-A.9). Off-diagonal matrix elements that induce nuclear flip-flops are highlighted in yellow.

| | $\|\uparrow\uparrow\uparrow\rangle$ | $\|\uparrow\uparrow\downarrow\rangle$ | $\|\downarrow\uparrow\uparrow\rangle$ | $\|\downarrow\uparrow\downarrow\rangle$ | $\|\uparrow\downarrow\uparrow\rangle$ | $\|\uparrow\downarrow\downarrow\rangle$ | $\|\downarrow\downarrow\uparrow\rangle$ | $\|\downarrow\downarrow\downarrow\rangle$ |
|---|---|---|---|---|---|---|---|---|
| $\langle\uparrow\uparrow\uparrow\|$ | $\frac{\omega_e}{2}-\omega_I+\frac{A_1^{zz}+A_2^{zz}}{4}$ | $A_2^{zx}/4$ | $A_1^{zx}/4$ | 0 | 0 | 0 | 0 | 0 |
| $\langle\uparrow\uparrow\downarrow\|$ | $A_{2s}^{zx}/4$ | $\frac{\omega_e}{2}+\frac{A_1^{zz}-A_2^{zz}}{4}$ | 0 | $A_1^{zx}/4$ | $A_2^{\perp}/2$ | 0 | 0 | 0 |
| $\langle\downarrow\uparrow\uparrow\|$ | $A_1^{zx}/4$ | 0 | $\frac{\omega_e}{2}+\frac{-A_1^{zz}+A_2^{zz}}{4}$ | $A_2^{zx}/4$ | $A_1^{\perp}/2$ | 0 | 0 | 0 |
| $\langle\downarrow\uparrow\downarrow\|$ | 0 | $A_1^{zx}/4$ | $A_2^{zx}/4$ | $\frac{\omega_e}{2}+\omega_I-\frac{A_1^{zz}+A_2^{zz}}{4}$ | 0 | $A_1^{\perp}/2$ | $A_2^{\perp}/2$ | 0 |
| $\langle\uparrow\downarrow\uparrow\|$ | 0 | $A_2^{\perp}/2$ | $A_1^{\perp}/2$ | 0 | $-\frac{\omega_e}{2}-\omega_I-\frac{A_1^{zz}+A_2^{zz}}{4}$ | $-A_2^{zx}/4$ | $-A_1^{zx}/4$ | 0 |
| $\langle\uparrow\downarrow\downarrow\|$ | 0 | 0 | 0 | $A_1^{\perp}/2$ | $-A_2^{zx}/4$ | $-\frac{\omega_e}{2}+\frac{-A_1^{zz}+A_2^{zz}}{4}$ | 0 | $-A_1^{zx}/4$ |
| $\langle\downarrow\downarrow\uparrow\|$ | 0 | 0 | 0 | $A_2^{\perp}/2$ | $-A_1^{zx}/4$ | 0 | $-\frac{\omega_e}{2}+\frac{A_1^{zz}-A_2^{zz}}{4}$ | $-A_2^{zx}/4$ |
| $\langle\downarrow\downarrow\downarrow\|$ | 0 | 0 | 0 | 0 | 0 | $-A_1^{zx}/4$ | $-A_2^{zx}/4$ | $-\frac{\omega_e}{2}+\omega_I+\frac{A_1^{zz}+A_2^{zz}}{4}$ |

**Table S3**. **Matrix representation for the Hamiltonian $H_a$.** Blue-shaded blocks correspond to different electronic Zeeman energy, which is the leading energy scale of the system.

| | $\lvert\uparrow'\uparrow\uparrow'\rangle$ | $\lvert\uparrow'\uparrow\downarrow'\rangle$ | $\lvert\downarrow'\uparrow\uparrow'\rangle$ | $\lvert\downarrow'\uparrow\downarrow'\rangle$ | $\lvert\uparrow'\downarrow\uparrow'\rangle$ | $\lvert\uparrow'\downarrow\downarrow'\rangle$ | $\lvert\downarrow'\downarrow\uparrow'\rangle$ | $\lvert\downarrow'\downarrow\downarrow'\rangle$ |
|---|---|---|---|---|---|---|---|---|
| $\langle\uparrow'\uparrow\uparrow'\rvert$ | $\frac{\omega_e}{2}+\frac{\Delta_1^{\uparrow}}{4}+\frac{\Delta_2^{\uparrow}}{4}$ | 0 | 0 | 0 | $\frac{A_1^{\perp}}{2}s_1^{\uparrow}c_1^{\downarrow}c_2+\frac{A_2^{\perp}}{2}s_2^{\uparrow}c_2^{\downarrow}c_1$ | $-\frac{A_1^{\perp}}{2}s_1^{\uparrow}c_1^{\downarrow}s_2-\frac{A_2^{\perp}}{2}s_2^{\uparrow}s_2^{\downarrow}c_1$ | $-\frac{A_1^{\perp}}{2}s_1^{\uparrow}s_1^{\downarrow}c_2-\frac{A_2^{\perp}}{2}s_2^{\uparrow}c_2^{\downarrow}s_1$ | $\frac{A_1^{\perp}}{2}s_1^{\uparrow}s_1^{\downarrow}s_2+\frac{A_2^{\perp}}{2}s_2^{\uparrow}s_2^{\downarrow}s_1$ |
| $\langle\uparrow'\uparrow\downarrow'\rvert$ | 0 | $\frac{\omega_e}{2}+\frac{\Delta_1^{\uparrow}}{4}-\frac{\Delta_2^{\uparrow}}{4}$ | 0 | 0 | $\frac{A_1^{\perp}}{2}s_1^{\uparrow}c_1^{\downarrow}s_2+\frac{A_2^{\perp}}{2}c_2^{\uparrow}c_2^{\downarrow}c_1$ | $\frac{A_1^{\perp}}{2}s_1^{\uparrow}c_1^{\downarrow}c_2-\frac{A_2^{\perp}}{2}c_2^{\uparrow}s_2^{\downarrow}c_1$ | $-\frac{A_1^{\perp}}{2}s_1^{\uparrow}s_1^{\downarrow}s_2-\frac{A_2^{\perp}}{2}c_2^{\uparrow}c_2^{\downarrow}s_1$ | $-\frac{A_1^{\perp}}{2}s_1^{\uparrow}s_1^{\downarrow}c_2+\frac{A_2^{\perp}}{2}c_2^{\uparrow}s_2^{\downarrow}s_1$ |
| $\langle\downarrow'\uparrow\uparrow'\rvert$ | 0 | 0 | $\frac{\omega_e}{2}-\frac{\Delta_1^{\uparrow}}{4}+\frac{\Delta_2^{\uparrow}}{4}$ | 0 | $\frac{A_1^{\perp}}{2}c_1^{\uparrow}c_1^{\downarrow}c_2+\frac{A_2^{\perp}}{2}s_2^{\uparrow}c_2^{\downarrow}s_1$ | $-\frac{A_1^{\perp}}{2}c_1^{\uparrow}c_1^{\downarrow}s_2-\frac{A_2^{\perp}}{2}s_2^{\uparrow}s_2^{\downarrow}s_1$ | $-\frac{A_1^{\perp}}{2}c_1^{\uparrow}s_1^{\downarrow}c_2+\frac{A_2^{\perp}}{2}s_2^{\uparrow}c_2^{\downarrow}c_1$ | $\frac{A_1^{\perp}}{2}c_1^{\uparrow}s_1^{\downarrow}s_2-\frac{A_2^{\perp}}{2}s_2^{\uparrow}s_2^{\downarrow}c_1$ |
| $\langle\downarrow'\uparrow\downarrow'\rvert$ | 0 | 0 | 0 | $\frac{\omega_e}{2}-\frac{\Delta_1^{\uparrow}}{4}-\frac{\Delta_2^{\uparrow}}{4}$ | $\frac{A_1^{\perp}}{2}c_1^{\uparrow}c_1^{\downarrow}s_2+\frac{A_2^{\perp}}{2}c_2^{\uparrow}c_2^{\downarrow}s_1$ | $\frac{A_1^{\perp}}{2}c_1^{\uparrow}c_1^{\downarrow}c_2-\frac{A_2^{\perp}}{2}c_2^{\uparrow}s_2^{\downarrow}s_1$ | $-\frac{A_1^{\perp}}{2}c_1^{\uparrow}s_1^{\downarrow}s_2+\frac{A_2^{\perp}}{2}c_2^{\uparrow}c_2^{\downarrow}c_1$ | $-\frac{A_1^{\perp}}{2}c_1^{\uparrow}s_1^{\downarrow}c_2-\frac{A_2^{\perp}}{2}c_2^{\uparrow}s_2^{\downarrow}c_1$ |
| $\langle\uparrow'\downarrow\uparrow'\rvert$ | $\frac{A_1^{\perp}}{2}s_1^{\uparrow}c_1^{\downarrow}c_2+\frac{A_2^{\perp}}{2}s_2^{\uparrow}c_2^{\downarrow}c_1$ | $\frac{A_1^{\perp}}{2}s_1^{\uparrow}c_1^{\downarrow}s_2+\frac{A_2^{\perp}}{2}c_2^{\uparrow}c_2^{\downarrow}c_1$ | $\frac{A_1^{\perp}}{2}c_1^{\uparrow}c_1^{\downarrow}c_2+\frac{A_2^{\perp}}{2}s_2^{\uparrow}c_2^{\downarrow}s_1$ | $\frac{A_1^{\perp}}{2}c_1^{\uparrow}c_1^{\downarrow}s_2+\frac{A_2^{\perp}}{2}c_2^{\uparrow}c_2^{\downarrow}s_1$ | $-\frac{\omega_e}{2}-\frac{\Delta_1^{\downarrow}}{4}-\frac{\Delta_2^{\downarrow}}{4}$ | 0 | 0 | 0 |
| $\langle\uparrow'\downarrow\downarrow'\rvert$ | $-\frac{A_1^{\perp}}{2}s_1^{\uparrow}c_1^{\downarrow}s_2-\frac{A_2^{\perp}}{2}s_2^{\uparrow}s_2^{\downarrow}c_1$ | $-\frac{A_1^{\perp}}{2}s_1^{\uparrow}c_1^{\downarrow}c_2-\frac{A_2^{\perp}}{2}c_2^{\uparrow}s_2^{\downarrow}c_1$ | $-\frac{A_1^{\perp}}{2}c_1^{\uparrow}c_1^{\downarrow}s_2-\frac{A_2^{\perp}}{2}s_2^{\uparrow}s_2^{\downarrow}s_1$ | $\frac{A_1^{\perp}}{2}c_1^{\uparrow}c_1^{\downarrow}c_2-\frac{A_2^{\perp}}{2}c_2^{\uparrow}s_2^{\downarrow}s_1$ | 0 | $-\frac{\omega_e}{2}-\frac{\Delta_1^{\downarrow}}{4}+\frac{\Delta_2^{\downarrow}}{4}$ | 0 | 0 |
| $\langle\downarrow'\downarrow\uparrow'\rvert$ | $-\frac{A_1^{\perp}}{2}s_1^{\uparrow}s_1^{\downarrow}c_2-\frac{A_2^{\perp}}{2}s_2^{\uparrow}c_2^{\downarrow}s_1$ | $-\frac{A_1^{\perp}}{2}s_1^{\uparrow}s_1^{\downarrow}s_2-\frac{A_2^{\perp}}{2}c_2^{\uparrow}c_2^{\downarrow}s_1$ | $-\frac{A_1^{\perp}}{2}c_1^{\uparrow}s_1^{\downarrow}c_2+\frac{A_2^{\perp}}{2}s_2^{\uparrow}c_2^{\downarrow}c_1$ | $-\frac{A_1^{\perp}}{2}c_1^{\uparrow}s_1^{\downarrow}s_2+\frac{A_2^{\perp}}{2}c_2^{\uparrow}c_2^{\downarrow}c_1$ | 0 | 0 | $-\frac{\omega_e}{2}+\frac{\Delta_1^{\downarrow}}{4}-\frac{\Delta_2^{\downarrow}}{4}$ | 0 |
| $\langle\downarrow'\downarrow\downarrow'\rvert$ | $\frac{A_1^{\perp}}{2}s_1^{\uparrow}s_1^{\downarrow}s_2+\frac{A_2^{\perp}}{2}s_2^{\uparrow}s_2^{\downarrow}s_1$ | $-\frac{A_1^{\perp}}{2}s_1^{\uparrow}s_1^{\downarrow}c_2+\frac{A_2^{\perp}}{2}c_2^{\uparrow}s_2^{\downarrow}s_1$ | $\frac{A_1^{\perp}}{2}c_1^{\uparrow}s_1^{\downarrow}s_2-\frac{A_2^{\perp}}{2}s_2^{\uparrow}s_2^{\downarrow}c_1$ | $-\frac{A_1^{\perp}}{2}c_1^{\uparrow}s_1^{\downarrow}c_2-\frac{A_2^{\perp}}{2}c_2^{\uparrow}s_2^{\downarrow}c_1$ | 0 | 0 | 0 | $-\frac{\omega_e}{2}+\frac{\Delta_1^{\downarrow}}{4}+\frac{\Delta_2^{\downarrow}}{4}$ |

**Table S4**. **Matrix representation for the Hamiltonian $\widetilde{H}_a$.** Here we follow and extend the notation introduced in Eqns. (A.4-A9), i.e. $\Delta_i^{\uparrow,\downarrow}=\Delta_i^{m_{S'}}= 2\sqrt{(m_{S'}A_i^{zx})^2+(m_{S'}A_i^{zz}-\omega_I)^2}$, $tan(\theta_i^{m_{S'}})=m_{S'}A_i^{zx}/(m_{S'}A_i^{zz}-\omega_I)$, $c_i^{\uparrow(\downarrow)}=\cos\left(\theta_i^{\uparrow(\downarrow)}/2\right)$, $s_i^{\uparrow(\downarrow)}=\sin\left(\theta_i^{\uparrow(\downarrow)}/2\right)$, $c_i=\cos\left(\frac{\theta_i^{\downarrow}-\theta_i^{\uparrow}}{2}\right)$ and $s_i=\sin\left(\frac{\theta_i^{\downarrow}-\theta_i^{\uparrow}}{2}\right)$.